\documentclass[pra, aps, twocolumn, floatfix, showpacs]{revtex4-1}
\usepackage{graphicx, amsmath, amssymb, times}

\begin{document}
\title{Attractive Hofstadter-Hubbard model with imbalanced chemical and vector potentials}
\author{M. Iskin}
\affiliation{
Department of Physics, Ko\c c University, Rumelifeneri Yolu, 34450 Sar{\i}yer, Istanbul, Turkey.
}
\date{\today}

\begin{abstract}

We study the interplay between the Hofstadter butterfly, strong interactions and
Zeeman field within the mean-field Bogoliubov-de Gennes theory in real space, 
and explore the ground states of the attractive single-band Hofstadter-Hubbard 
Hamiltonian on a square lattice, including the exotic possibility of imbalanced 
vector potentials. 
We find that the cooperation between the vector potential and superfluid 
order breaks the spatial symmetry of the system, and flourish stripe-ordered 
Fulde-Ferrell-Larkin-Ovchinnikov (FFLO)-like superfluid and supersolid 
phases that can be distinguished and 
characterized according to their coexisting pair-density (PDW), charge-density 
(CDW) and spin-density (SDW) wave orders. 
We also discuss confined systems and comment on the likelihood of observing 
such stripe-ordered phases by loading neutral atomic Fermi gases on 
laser-induced optical lattices under laser-generated artificial gauge fields.

\end{abstract}
\pacs{03.75.Ss, 03.75.Hh, 67.85.-Lm, 67.85.-d, 67.80.kb}
\maketitle

\section{Introduction}
\label{sec:intro}

The exact energy spectrum of a single quantum particle that is confined to 
move on a two-dimensional tight-binding periodic lattice under the influence 
of a uniform magnetic flux has been known for a long time~\cite{hofstadter76, kohmoto89},
where the competition between the lattice spacing and cyclotron radius
gives rise to a self-similar complex pattern of sub-bands and mini-gaps. 
However, regardless of all efforts since the prediction of this Hofstadter 
spectrum, there has been very recent but still limited success 
in observing some of its signatures in graphene-based solid-state materials 
with artificially-engineered superlattices under real magnetic 
fields~\cite{dean13, ponomarenko13}. In addition, thanks to the recent 
realisation of artificial gauge fields in atomic 
systems~\cite{galitski13, dalibard11, chen12, wang12,cheuk12,qu13,fu13, williams13}, 
there is also an increasing interest on this subject from the cold-atom 
community~\cite{garcia12, struck12, kennedy13, aidelsburger13, miyake13, chin13, cocks12, wang14}. 
In particular, by engineering spatially-dependent complex tunneling 
amplitudes with laser-assisted tunneling and a potential energy gradient, 
two groups have recently reported realisation of the Hofstadter-Harper 
Hamiltonian using neutral rubidium atoms that are loaded into laser-induced 
periodic potentials~\cite{aidelsburger13, miyake13, chin13}.  

Even though the Hofstadter and Hubbard Hamiltonians have themselves been 
the subject of many works in the literature, there has been a lack of interest in 
the combined Hofstadter-Hubbard Hamiltonian even at the mean-field level.
For instance, while the use of momentum-space BCS formalism limits previous
analysis of the attractive Hofstadter-Hubbard model only to vortex lattice 
(VL) configurations~\cite{zhai10}, the existence of pair-density wave (PDW) 
and VL orders have been proposed in the context of a 
somewhat related model: an anisotropic 3D continuum Fermi gas experiencing 
a uniform magnetic flux~\cite{wei12}. By first limiting their description to the 
lowest-Landau-level limit and then making further assumptions about the 
strength of the anisotropic trap, the authors obtain an effectively a 1D Hamiltonian 
in momentum space, and solved it using the BCS formalism. The existence 
and characterisation of a variety of distinct stripe-ordered many-body phases 
have either been overlooked or gone unnoticed until very 
recently~\cite{miskin-stripe}, distinguishing our work from the literature.

In particular, here we use Bogoliubov-de Gennes (BdG) theory in real space and 
study the mean-field ground states of the attractive single-band 
Hofstadter-Hubbard Hamiltonian on a square lattice, including the effects of 
imbalanced chemical and vector potentials. We find that the cooperation 
between the vector potentials and interaction breaks the spatial symmetry of 
the system, leading to various stripe-ordered superfluid (SF) and supersolid 
(SS) phases that can be distinguished and characterized according to their 
coexisting PDW, charge-density (CDW) and spin-density (SDW) wave orders. 
We also discuss possible observation 
of such stripe-ordered phases by confining neutral atomic Fermi gases in 
laser-induced optical lattices under laser-generated artificial gauge fields.

The rest of this paper is organised as follows. 
In Sec.~\ref{sec:theory}, first we introduce the physical setting of the 
problem and the model Hamiltonian used, then review the non-interacting 
Hofstadter Hamiltonian and its well-known Hofstadter spectrum, and then describe 
the self-consistent BdG formalism which takes fermion-fermion interactions 
into account within the mean-field approximation for pairing.
The resultant BdG equations are solved in Sec.~\ref{sec:numerics}, 
where first we tabulate the numerically obtained mean-field ground states, 
paying a special attention to the striped phases in the dimer-BEC limit, 
and then construct the thermodynamic phase diagrams. The effects of 
Hartree shifts on the possible ground states are discussed in Sec.~\ref{sec:cas} 
in the context of harmonically-confined atomic systems. 
We end the paper with a briery summary of our conclusions and an 
outlook in Sec.~\ref{sec:conc}, and an Appendix comparing the dimer-BEC limit 
in the Landau and symmetric gauges.

\section{Theoretical Framework}
\label{sec:theory}

To explore the ground states of the single-band Hofstadter-Hubbard model,
we start with
\begin{align}
\label{eqn:eham}
H = &- \sum_{ij\sigma} t_{ij \sigma} a_{i\sigma}^\dagger a_{j\sigma}
-\sum_{i\sigma} \mu_{i \sigma} a_{i\sigma}^\dagger a_{i\sigma} \nonumber \\
&- g \sum_{i} a_{i\uparrow}^\dagger a_{i\uparrow} a_{i\downarrow}^\dagger a_{i\downarrow},
\end{align}
and consider both thermodynamic and confined systems.
Here, $a_{i\sigma}^\dagger$ ($a_{i\sigma}$) creates (annihilates) a
$\sigma \equiv \lbrace \uparrow, \downarrow \rbrace$ fermion on site $i$, 
$t_{ij \sigma}$ is its hopping parameter from site $i$ to $j$, and 
$\mu_{i \uparrow} = \mu - V_i + h$ 
and
$\mu_{i \downarrow} = \mu - V_i - h$ 
are effectively their local chemical potentials in the presence of confining 
potential $V_i$ and an out-of-plane Zeeman field $h$. 
We assume $h \ge 0$ without loosing generality, since $h < 0$ results 
can be easily deduced by letting $\uparrow \to \downarrow$ 
and $\downarrow \to \uparrow$.
The density-density interaction term is taken to be local (on-site) and 
attractive with strength $g \ge 0$,  and the resultant many-body phases 
are investigated within the mean-field approximation for the Cooper pairs 
and their superfluidity, as described below.

\subsection{Mean-Field Hofstadter-Hubbard Hamiltonian}
\label{sec:ham}
In particular, we analyse the following mean-field Hamiltonian for square
lattices,
\begin{align}
\label{eqn:ham}
H_{mf} = &- \sum_{ij\sigma} t_{ij \sigma} a_{i\sigma}^\dagger a_{j\sigma}
-\sum_{i\sigma} \widetilde{\mu}_{i \sigma} a_{i\sigma}^\dagger a_{i\sigma} \nonumber \\
&+ \sum_{i} \left( \Delta_i a_{i\uparrow}^\dagger a_{i\downarrow}^\dagger 
+ \Delta_i^* a_{i\downarrow} a_{i\uparrow} + \frac{|\Delta_i|^2}{g} \right),
\end{align}
where 
$\widetilde{\mu}_{i \uparrow} = \mu_{i \uparrow} - gn_{i\downarrow}$ 
and
$\widetilde{\mu}_{i \downarrow} = \mu_{i \downarrow} - gn_{i\uparrow}$ 
take the Hartree shifts into account.
Here, $n_{i \sigma} = \langle a_{i\sigma}^\dagger a_{i\sigma} \rangle$ 
is the average number of $\sigma$ fermions where $\langle \cdots \rangle$ 
is a thermal average, and the remaining terms in Eq.~(\ref{eqn:ham}) 
involve the complex SF order parameter
$
\Delta_i = g \langle a_{i\uparrow} a_{i\downarrow} \rangle.
$
These average quantities are specified below in 
Eqs.~(\ref{eqn:op})-(\ref{eqn:ndo}), and we use them in Sec.~\ref{sec:numerics} 
for characterising the state of the system. 

The complex hopping parameters are assumed to connect only the 
nearest-neighbor sites, i.e, $t_{ij \sigma} = t_\sigma e^{i\theta_{ij \sigma}}$
where the amplitudes $t_\uparrow = t_\downarrow = t \ge 0$ are taken to be 
equal for $i$ and $j$ nearest neighbors and $0$ otherwise. 
The phase, however, depends on the external magnetic (or artificial gauge)
field experienced by $\sigma$ fermions. 
In particular, we use the Peierls substitution and take
$
\theta_{ij \sigma} = (1/\phi_0) \int_{\mathbf{r}_i}^{\mathbf{r}_j} 
\mathbf{A_\sigma}(\mathbf{r}) \cdot d\mathbf{r},
$
with $\phi_0 = 2\pi \hbar/e$ the magnetic flux quantum and 
$\mathbf{A_\sigma}(\mathbf{r})$ the vector potential which is assumed to 
be independently controllable for $\uparrow$ and $\downarrow$ fermions. 
Note that while independent control of $\mathbf{A_\sigma}(\mathbf{r})$ 
is not possible for conventional solid-state materials with real magnetic 
fields where $\sigma$ corresponds to the $\pm$ projections of spin 
angular momentum of electrons, such a control can be achieved with 
neutral atomic systems under the influence of laser-generated artificial 
gauge fields where pseudo-spin $\sigma$ is just a label for two of 
the hyperfine states of a particular atom. In this paper, we choose Landau 
gauge for the vector potential, i.e, 
$\mathbf{A_\sigma}(\mathbf{r}) \equiv (0,B_\sigma x,0)$, 
leading to a uniform magnetic flux $\Phi_\sigma = B_\sigma \ell^2$ per unit 
cell penetrating our square lattice, where $\ell$ is the lattice spacing.
Denoting ($x,y$) coordinates of site $i$ by ($n\ell, m\ell$), 
this gauge simply implies $\theta_{ij \sigma} = 0$ and 
$\theta_{ij \sigma} = \pm 2\pi n\phi_\sigma$ 
for links along the $x$ and $y$ directions, respectively,
where $\phi_\sigma = \Phi_\sigma/(2\pi \phi_0)$ characterizes the 
competition between $\ell$ and the magnetic length scale (cyclotron radius) 
$\ell_{B_\sigma} = \sqrt{\hbar/(eB_\sigma)}$. 
We note that while $\phi_\uparrow = \phi_\downarrow \ll 1$ for typical 
electronic crystals, even for the largest magnetic field 
$B_\uparrow = B_\downarrow \sim 100 T$ that is attainable in a laboratory, 
$\phi_\uparrow$ and $\phi_\downarrow$ may be tuned at will in 
atomic optical lattices. 

Let us first set $g = 0$ and $\mu_{i\sigma} = 0$ in Eq.~(\ref{eqn:ham}), and 
review the well-known single-particle problem, i.e, the Hofstadter 
Hamiltonian for a uniform square lattice.

\subsection{Hofstadter Butterfly (HB)}
\label{sec:hb}

In the non-interacting limit, the single-particle Hofstadter-Hamiltonian 
describing a $\sigma$ fermion can be written as,
\begin{align}
H_{0\sigma} = - t_\sigma \sum_{nm} &\bigg{(} a_{nm\sigma}^\dagger a_{n+1,m\sigma} \nonumber \\
& + e^{i 2\pi\phi_\sigma n} a_{nm\sigma}^\dagger a_{n,m+1,\sigma} + \textrm{H.c.}\bigg{)},
\label{eqn:h0}
\end{align}
where H.c. is the Hermitian conjugate. 
For rational values of $\phi_\sigma \equiv p_\sigma/q_\sigma$, where $p_\sigma$ 
and $q_\sigma$ are positive integers with no-common factor, 
i.e, co-prime numbers, while $H_{0 \sigma}$ maintains its 
translational invariance in the $y$ direction, it requires $q_\sigma$ sites 
for translational invariance in the $x$ direction. Thanks to the Bloch 
theorem, the 1st magnetic Brillouin zone is determined by
$
-\pi \le k_y \ell \le \pi
$
and
$
-\pi/q_\sigma \le k_x \ell \le \pi/q_\sigma,
$
and this increased periodicity motivates us to work with a supercell of 
$1 \times q_\sigma$ sites. The excitation spectrum is determined 
by solving the Schr\"odinger equation 
$
H_{0 \sigma} \Psi_\sigma = \varepsilon (\phi_\sigma) \Psi_\sigma
$ 
for all momentum $\mathbf{k} \equiv (k_x, k_y)$ values in the 1st magnetic 
Brillouin zone. Denoting the components of the wave function as
$
\Psi_{\sigma} = ( \psi_1^*, \psi_2^*, \psi_3^*, \dots, \psi_{q_\sigma}^*)^\dagger,
$
where $\psi_n$ corresponds to the $n$th site of the supercell, 
the $q_\sigma \times q_\sigma$ Hamiltonian matrix at a given $\mathbf{k}$ value 
%
\begin{equation}
\label{eqn:hbmatrix}
\begin{bmatrix}
C_{1\sigma} & T_\sigma^* & 0 & . & . & 0 & T_\sigma \\
T_\sigma & C_{2\sigma} & T_\sigma^* & 0 & . & . & 0 \\
0 & T_\sigma & C_{3\sigma} & . & . & . & . \\
. & . & . & . & . & . & 0 \\
0 & . & . & . & . & C_{n-1,\sigma} & T_\sigma^* \\
T_\sigma^* & 0 & . & . & 0 & T_\sigma & C_{n\sigma} \\
\end{bmatrix}
\end{equation}
describes the supercell with periodic Bloch boundary conditions. 
Here, $C_{n\sigma} = -2 t_\sigma \cos(k_y\ell + 2\pi n p_\sigma/q_\sigma)$ 
and $T_\sigma = -t_\sigma e^{ik_x\ell}$.

\begin{figure}[htb]
\centerline{\scalebox{0.4}{\includegraphics{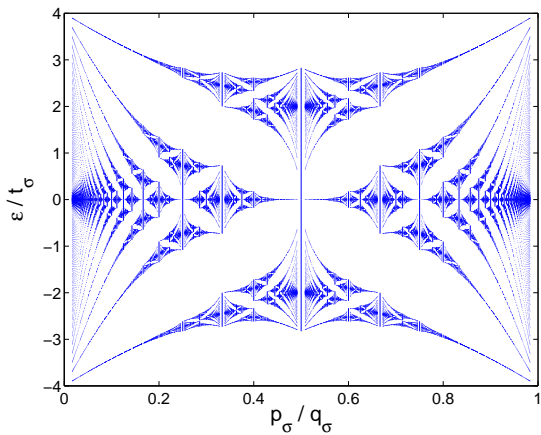}}}
\caption{\label{fig:hb} (Color online)
The Hofstadter spectrum $\varepsilon/t_\sigma$ is presented 
as a function of $\phi_\sigma =  p_\sigma/q_\sigma$, showing its fractal 
structure with numerous sub-bands and mini-gaps.
}
\end{figure}

The eigenvalues $\varepsilon(\phi_\sigma)$ of this $q_\sigma \times q_\sigma$ 
matrix can be numerically obtained for any given rational number $\phi_\sigma$
and the energy spectrum $\varepsilon(\phi_\sigma)$ vs. $\phi$ is known as 
HB~\cite{hofstadter76, kohmoto89}. 
The spectrum is shown in Fig.~\ref{fig:hb}, where, for a given $\phi_\sigma$, it 
consists of non-overlapping $q_\sigma$ bands with $q_\sigma+1$ energy gaps 
in between, and each one of these $q_\sigma$ bands can accommodate 
$1/q_\sigma$ particle filling with a total filling of $1$. Therefore, if we index energy 
gaps as $z_\sigma = \lbrace 0, 1, 2, \cdots, q_\sigma \rbrace$, 
starting from the bottom edge of the band in such a way that the lowest 
($z_\sigma = 0$) and highest ($z_\sigma = q_\sigma$) gaps correspond, 
respectively, to a particle vacuum and a fully-filled band insulator, particle fillings 
within all of these gapped regions can be compactly written as $z_\sigma/q_\sigma$. 
Note that while all gaps are open for odd $q_\sigma$, the middle 
$z_\sigma = q_\sigma/2$ gap corresponding to a half-filled lattice is not open 
when $q_\sigma$ is even, and therefore, a half-filled lattice is not an insulator 
for any $q_\sigma$. In the $\phi_\sigma \to 0$ limit, the HB spectrum recovers 
the usual tight-binding dispersion of cosines 
$
\varepsilon_{\mathbf{k} \sigma} = -2t_\sigma \left[\cos(k_x \ell) + \cos(k_y \ell) \right],
$
which has an energy bandwidth $W_\sigma = 8t_\sigma$.

Since $\ell$ and $\ell_{B_\sigma}$ are the only two length scales in 
Eq.~(\ref{eqn:h0}) such that $\phi_\sigma = \ell^2/(2 \pi \ell_{B_\sigma}^2)$, 
the fractal structure of HB is clearly a result of their competition. 
In addition, each $\mathbf{k}$ state is $q_\sigma$-fold degenerate in the 1st 
magnetic Brillouin zone (not explicitly shown in the figure), i.e, 
\begin{align}
\label{eqn:hbe}
\varepsilon_{b, k_x k_y} (\phi_\sigma) = \varepsilon_{b, k_x, k_y + 2\pi \phi_\sigma f/\ell} (\phi_\sigma)
\end{align}
with $b = 1,2,\cdots,q_\sigma$ labelling the bands and $f = 1,2,\cdots,q_\sigma$ 
labelling the degenerate $\mathbf{k}$ states. We have recently shown 
that the HB spectrum plays a crucial role in determining the many-body 
states of the interacting system~\cite{miskin-stripe}, and our primary objective 
here is to extend and generalise the analysis to imbalanced gauge fields.

\subsection{Bogoliubov-de Gennes (BdG) Theory}
\label{sec:bdg}
For this purpose, we diagonalise Eq.~(\ref{eqn:ham}) via the Bogoliubov-Valatin 
transformation, i.e,
$
a_{i\sigma} = \sum_m (u_{mi\sigma} \gamma_{m\sigma} 
- s_\sigma v_{mi\sigma}^* \gamma_{m,-\sigma}^\dagger),
$
where $\gamma_{m\sigma}^\dagger$ ($\gamma_{m\sigma}$) creates (annihilates) 
a pseudo-spin $\sigma$ quasiparticle with energy $\epsilon_{m}^\sigma$ and wave 
functions $u_{mi\sigma}$ and $v_{mi\sigma}$, and $s_\uparrow = +1$ and 
$s_\downarrow = -1$. The resultant BdG equations can be compactly written as,
\begin{equation}
\label{eqn:bdg}
\sum_j 
\left( \begin{array}{cc}
-t_{ij\uparrow} - \widetilde{\mu}_{i\uparrow} \delta_{ij} & \Delta_i \delta_{ij} \\
\Delta_i^*\delta_{ij} & t_{ij\downarrow}^* + \widetilde{\mu}_{i\downarrow} \delta_{ij}
\end{array} \right)
\varphi_{mj}^\sigma
 = s_\sigma \epsilon_{m}^\sigma \varphi_{mi}^\sigma,
\end{equation}
where $\delta_{ij}$ is the Kronecker delta, and
$
\varphi_{mi}^{\uparrow} = (u_{mi\uparrow}^*, v_{mi\downarrow}^*)^\dagger
$
and
$
\varphi_{mi}^{\downarrow} = (v_{mi\uparrow}, -u_{mi\downarrow})^\dagger
$
are the corresponding eigenfunctions for $\epsilon_{m}^\sigma \ge 0$ eigenvalue.
Note that the BdG equations are invariant under the transformation
$v_{mi\uparrow} \to u_{mi\uparrow}^*$, $u_{mi\downarrow} \to -v_{mi\downarrow}^*$ 
and $\epsilon_{m\downarrow} \to -\epsilon_{m\uparrow}$, and therefore, it is 
sufficient to solve only for $u_{mi} \equiv u_{mi\uparrow}$, $v_{mi} \equiv v_{mi\downarrow}$ 
and $\epsilon_m \equiv \epsilon_{m}^\uparrow$ as long as all solutions with 
positive and negative $\epsilon_m$ are kept.

Using the transformation, the complex order parameter $\Delta_i$ can be 
written as
\begin{align}
\label{eqn:op}
\Delta_i = - g\sum_m u_{mi} v_{mi}^* f(\epsilon_m),
\end{align}
where $f(x) = 1/[e^{x/(k_B T)} + 1]$ is the Fermi function with $k_B$ the 
Boltzmann constant and $T$ the temperature. 
Equations~(\ref{eqn:bdg}) and~(\ref{eqn:op}) have to be solved self-consistently 
for a given $\mu$ and $h$, such that the total number of $\sigma$ fermions 
satisfies $N_\sigma = \sum_i n_{i\sigma}$. Here,
$
0 \le n_{i \sigma} = \langle a_{i\sigma}^\dagger a_{i\sigma} \rangle \le 1
$
is the average number of $\sigma$ fermions on site $i$, and using the 
transformation, it can be written as
\begin{align}
\label{eqn:nup}
n_{i\uparrow} = \sum_m |u_{mi}|^2 f(\epsilon_m), \\
\label{eqn:ndo}
n_{i\downarrow} = \sum_m |v_{mi}|^2 f(-\epsilon_m),
\end{align}
for the $\uparrow$ and $\downarrow$ fermions, respectively.
We note that unlike the continuum models where the solutions of the
self-consistency equations depend explicitly on the high-momentum 
cut-off, requiring a high-energy regularisation in order to obtain cut-off 
independent results, the lattice versions given in 
Eqs.~(\ref{eqn:op})-(\ref{eqn:ndo}) do not require such a regularisation, 
since the lattice spacing $\ell$ already provides an implicit short-distance 
cut-off.

In the absence of gauge fields when $\theta_{ij} = 0$, it is generally 
accepted that the mean-field description given above provides qualitative 
understanding either at low temperatures ($T \ll T_c$) for any $g$ or 
for weak $g \lesssim W$ at any $T$, where $T_c$ is the critical SF transition 
temperature. It is also known that single-band Hubbard models gradually 
become inadequate in describing strongly-interacting cold-atom systems
on optical lattices, requiring multi-band models~\cite{fhreview}. In addition, 
the real-space BdG theory goes beyond the standard local-density 
approximation since it includes both $\theta_{ij}$ and $V_i$ exactly into 
the mean-field theory without relying on further approximations. 
Hoping to shed light on the qualitative effects of gauge fields on the 
ground states of Eq.~(\ref{eqn:ham}), here we mainly concentrate on weak 
and intermediate $g$ at $T = 0$ as discussed next.

\section{Numerical Framework}
\label{sec:numerics}

In order to explore the possible phases, let us set $V_i = 0$ and consider a 
uniform $45 \ell \times 45 \ell$ square lattice, which is large enough to construct the 
thermodynamic phase diagrams for $\phi_\sigma = \lbrace 0, 1/6, \pm 1/4 \rbrace$. 
We note that even though our phase diagrams are reliable, the phase boundaries 
should be taken as qualitative guides to the eye due to the possibility of minor 
finite-size effects.
We neglect the Hartree shifts for the moment because not only the 
self-consistent solutions converge much faster but also the resultant
phase diagrams are much more easier to interpret and understand. 
In addition, since none of the PDW, CDW and SDW instabilities are driven 
by these shifts, our qualitative mean-field results already paves the way to 
quantitative understanding of the possible ground states of Eq.~(\ref{eqn:ham}).
However, see Sec.~\ref{sec:hs} for the effects of Hartree shifts on 
confined systems.

For this purpose, we numerically solve Eqs.~(\ref{eqn:bdg})-(\ref{eqn:ndo})
at $T = 0$, and obtain self-consistent solutions of $\Delta_i/t$ and 
$n_{i\sigma}$ as functions of $g/t$, $\mu/t$, $h/t$ and $\phi_\sigma$. 
This can be achieved numerically via the iterative method of relaxation 
as follows. For a given set of parameters, first (i) start with an input set 
of $\Delta_i$, then (ii) construct the BdG matrix given in Eq.~(\ref{eqn:bdg}), 
and then (iii) use its eigenstates in Eq.~(\ref{eqn:op}) to 
generate a new set of $\Delta_i$, and finally (iv) repeat these steps until 
the input and output sets of $\Delta_i$ lie within a confidence level.
Once this iterative method converges, (v) use Eqs.~(\ref{eqn:nup})-(\ref{eqn:ndo}) 
to calculate $n_{i\sigma}$. It turns out that while Eqs.~(\ref{eqn:bdg})-(\ref{eqn:ndo}) 
have unique solutions in the low-$h/g$ limit, they in general allow for 
multiple solutions for the polarized many-body phases, and therefore, 
it is essential to try several initial sets of $\Delta_i$ and verify the 
(meta)stability of the solutions.

\begin{table}
\caption{\label{tab:sf}
While the S-SS* phase has a small but finite sign-changing striped-SDW 
order, the system is globally unpolarized very much like the unpolarized 
uniform superfluid (U-SF) or unpolarized striped supersolid (S-SS) 
phase. 
}
\begin{tabular}[c]{cccccc}
\hline
Phase & $|\Delta_i|$ & $n_{i\uparrow}+n_{i\downarrow}$ & $n_{i\uparrow}-n_{i\downarrow}$ & $\phi_\sigma$ \\
\hline
U-SF & Uniform & Uniform & 0 & $\phi_\uparrow = - \phi_\downarrow$ \\
S-SF & PDW & 1 & 0 & $\phi_\uparrow = \phi_\downarrow$ \\
S-SS & PDW & CDW & 0 & $\phi_\uparrow = \phi_\downarrow$ \\
S-SS* & PDW & CDW & SDW & $|\phi_\uparrow| \ne |\phi_\downarrow|$ \\
P-SF & \multicolumn{4} {c} {otherwise} \\
\hline
\end{tabular}
\end{table}
\subsection{Ground-State Phases}
\label{sec:gs}

Depending on the spatial profiles of $|\Delta_i|$, $n_{i\uparrow}$ and 
$n_{i\downarrow}$, we distinguish the single-particle band insulator and 
normal phases from the ordered many-body ones using the following criteria. 
When $h/g$ is sufficiently high that $\Delta_i \to 0$ (precisely speaking
$|\Delta_i| < 10^{-3}t$ in our numerics) for every $i$, the ground state 
can be a $\sigma$-vac phase which is a vacuum of $\sigma$ component 
with $n_{i\sigma} = 0$, a $\sigma$-$I(m/n)$ phase which is a band 
insulator of $\sigma$ component with uniform $n_{i\sigma} = m/n$,
a $\sigma$-$N$ phase which is a normal $\sigma$ component,
or an $\uparrow \downarrow$-$PN$ phase which is a polarized normal 
mixture of $\uparrow$ and $\downarrow$ components. 
We checked in our numerics that while $\sigma$-$N$ and $\uparrow \downarrow$-$PN$ 
phases have slightly non-uniform $n_{i\sigma}$ for $\phi \ne 0$, 
the $C_4$ symmetry of the square lattice is preserved.
On the other hand, when $h/g$ is sufficiently low that $\Delta_i \ne 0$
(i.e, $|\Delta_i| > 10^{-3}t$) for some $i$, the ground states 
can be characterized according to Table~\ref{tab:sf}.
Unlike our earlier work~\cite{miskin-stripe}, here we do not finely classify 
the polarized superfluid (P-SF) phase depending on the coexisting (striped or 
non-striped) PDW, CDW, SDW and/or VL orders. Instead, we focus mostly on 
the existence of striped phases in the dimer-BEC limit as the main message of 
this manuscript, for which physical (analytical) insight are also given.

\begin{figure}[htb]
\centerline{\scalebox{0.59}{\includegraphics{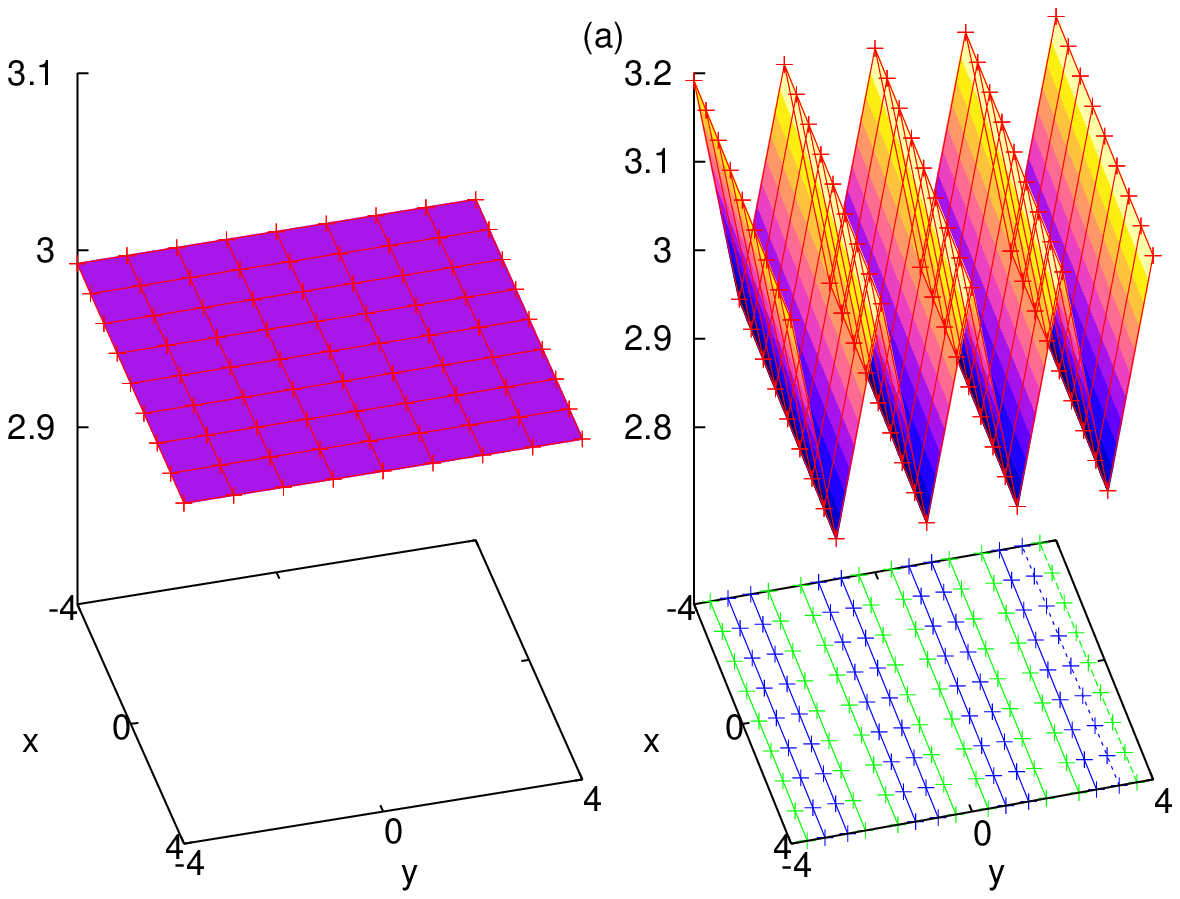}}}
\centerline{\scalebox{0.59}{\includegraphics{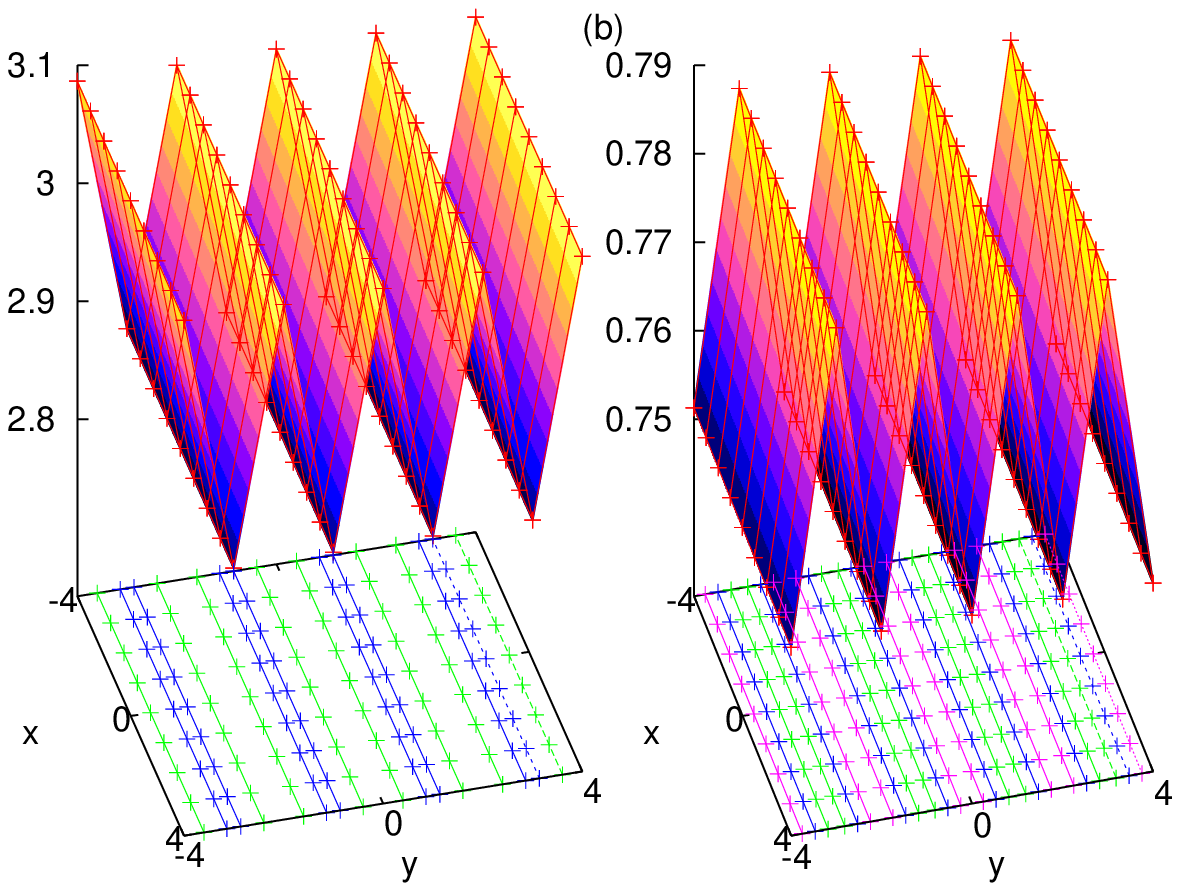}}}
\centerline{\scalebox{0.59}{\includegraphics{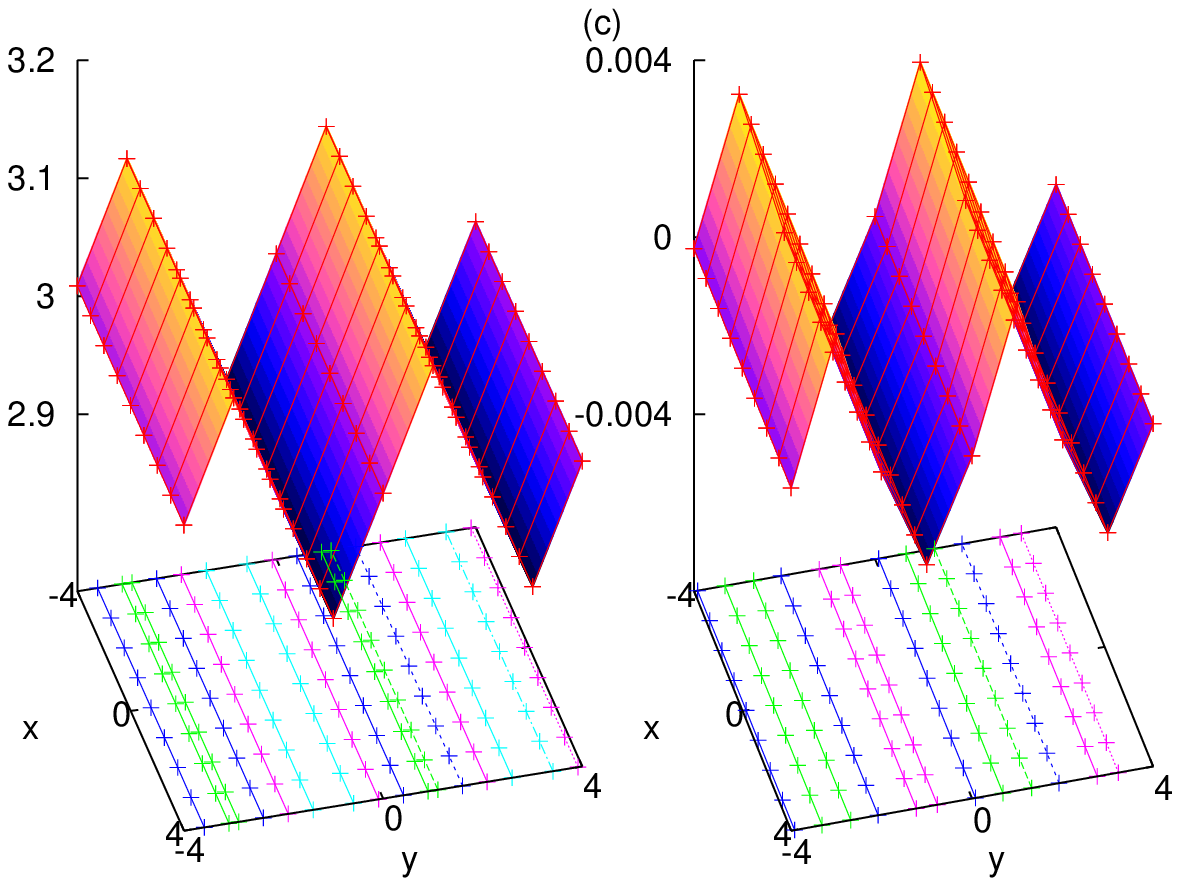}}}
\caption{\label{fig:stripes} (Color online) Characterisation of globally-unpolarized
many-body phases.
(a) Typical $|\Delta_i|/t$ profiles are shown for the U-SF (left) and 
S-SF phases (right), where $\phi_\uparrow = - \phi_\downarrow = 1/4$
and $\phi_\uparrow = \phi_\downarrow = 1/4$, respectively, and 
$\mu = 0$ (uniformly half-filled) in both figures.
(b) Typical $|\Delta_i|/t$ (left) and $n_{i\uparrow} + n_{i\downarrow}$ (right) 
profiles are shown for the S-SS phase, 
where $\phi_\uparrow = \phi_\downarrow = 1/4$ and $\mu = -t$.
(c) Typical $|\Delta_i|/t$ and $n_{i\uparrow} - n_{i\downarrow}$ profiles 
are shown for the S-SS* phase, 
where $\phi_\uparrow = 0$, $\phi_\downarrow = 1/4$ and $\mu = -t$. 
Note in (c) that even though the system is globally unpolarized, 
it has both SDW and CDW (not shown) orders.
Here, $(x,y)$ are in units of $\ell$, and we set $h = 0$ and $g = 7t$ in 
all figures.
}
\end{figure}

The globally-unpolarized states are denoted by U-SF, S-SF and S-SS,
and they stand, respectively, for uniform-SF, striped-SF, 
and striped-SS. The S-SS* state is also globally unpolarized, 
very much like the S-SS phase but it has an additional sign-changing 
striped-SDW order driven by the imbalance between $\phi_\uparrow$ 
and $\phi_\downarrow$. For instance, typical $|\Delta_i|$ and 
$n_{i\uparrow} \pm n_{i\downarrow}$ profiles are illustrated in Fig.~\ref{fig:stripes}
for all of them. Depending on $\mu$, $h$, $\phi_\uparrow$ and 
$\phi_\downarrow$, one of the U-SF, S-SF, S-SS and S-SS* phases 
always appears in the thermodynamic phase diagrams beyond a critical 
$g/t$ threshold, as discussed next.

\subsection{Dimer-BEC Limit in the Landau Gauge}
\label{sec:dimer}

When $g/t \gg 1$ is sufficiently high, the physics must eventually be determined 
by the two-body bound states, i.e, Cooper pairs become bosonic 
dimers, and unless $g/t \to \infty$, the dimer-dimer interaction 
[$g_{dd} \sim (t_\uparrow^2 +  t_\downarrow^2) /g$] 
is finite. Such weakly-repulsive dimers can effectively be described by the 
Hofstadter-Bose-Hubbard model, where superfluidity has recently been 
shown to break translation symmetry in the weakly-interacting 
limit~\cite{powell10}. 

In the ideal-dimer limit of our model Hamiltonian, the only way a tightly-bound
dimer to move from a site $i$ to $j$ in the lattice is via what is known as 
pair-breaking mechanism, i.e, virtual ionisation of its constituents 
costs a penalty of $g$, and this  gives rise to the effective dimer hopping 
parameter $t_{ijd} = 2 t_{ij \uparrow} t_{ij\downarrow}/g$. Therefore, the 
effective hopping amplitude and gauge field of the dimers can be written as 
$t_d \approx 2t_\uparrow t_\downarrow/g$ 
and $\phi_d = \phi_\uparrow + \phi_\downarrow = p_d/q_d$, respectively, 
where $p_d = (p_\uparrow q_\downarrow + p_\downarrow q_\uparrow)/Q$ 
and $q_d = q_\uparrow q_\downarrow/Q$. Here, $Q$ is a positive integer 
number chosen such that $p_d$ and $q_d$ are co-prime numbers,
and it depends on the entire
$\lbrace p_\uparrow, p_\downarrow, q_\uparrow, q_\downarrow \rbrace$ set.
Since HB for dimers is $q_d$-fold degenerate, their BEC order parameter 
has contributions from all degenerate 
$\mathbf{k_d} = \{ (0,0); (0,  2\pi \phi_d f / \ell) \}$ momenta, 
where $f = 1, \cdots, q_d-1$ such that
$
\Psi_{id} = c_0 + \sum_{f} c_f e^{i2\pi \phi_d f i_y / \ell}
$
and $c_f = |c_f| e^{i\vartheta_f}$ are complex variational parameters. 
However, unlike atomic bosons where all of the degenerate states have 
equal weight, dimer bosons are fermion pairs and the number of ways of 
creating them with $k_{yd} = k_{y \uparrow} + k_{y \downarrow}$ 
momentum depends on $f$, $\phi_\uparrow$ and $\phi_\downarrow$. 
For instance, there are $2(q-f)-1$ ways of intra-band pairing when 
$\phi_\uparrow = \phi_\downarrow = p/q$ and $q$ is even.
Thus, this analysis show that higher $k_{yd}$ states contribute less 
and less, forming a perturbative series. 

It turns out that the first order ($f = 1$) correction is already much smaller 
than the zeroth order ($f = 0$) one, and that the $f \ge 2$ terms are 
always negligible when $g/t$ is sufficiently large. This is because all of 
our numerical results fit quite well with 
\begin{align}
\label{eqn:fit}
|\Delta_i| = |\Delta_0| + |\Delta_1| \left[1 - \cos(2\pi \phi_d i_y/\ell) \right],
\end{align}
in the entire globally-unpolarized region, including S-SF, S-SS and 
S-SS* phases. 
Here, the $\mathbf{k_d} = (0,0)$ contribution $|\Delta_0| = (g/2-4t^2/g)\sqrt{n(2-n)}$ 
is uniform in space and determined by the total average filling $n$ with
$
\mu = (g/2-8t^2/g) (n-1)
$ 
~\cite{iskin-lattice}, 
$|\Delta_1| \approx t^2/g$ for $\mu \approx 0$, 
and $i_y$ is the $y$ coordinate of site $i$.
Moving towards the BCS side, the second-order correction to Eq.~(\ref{eqn:fit}) 
can be shown to be $+ |\Delta_2| \cos(4\pi \phi_d i_y/\ell)$ for even $q_d$.
Since this term is in- (out-of-$\pi$-) phase with the zeroth (first) order term, 
it tends to open throughs along the peaks arised from the first order one,
suggesting that $\vartheta_f - \vartheta_0 = \pi f$, i.e., the form of $|\Psi_{id}|$ 
coincides with $|\Delta_i|$ under these conditions.
Equation~(\ref{eqn:fit}) clearly shows that modulations of $|\Delta_i|$ 
have a spatial period of $q_d$ lattice sites along the $y$ direction.
It also implies that it is the cooperation between 
$\phi_d$ and $g$ that is responsible for the broken spatial symmetry 
and appearance of stripe order, and even though the stripe order gradually 
fades away with increasing $g$, it survives even in the $g \gg W$ limit 
as long as $g/t$ is finite. 

Thus, this analysis suggests that the existence of stripe-ordered SF and SS 
phases is not an artefact of the mean-field description, and they are 
physically expected in the dimer-BEC limit of the attractive Hofstadter-Hubbard 
model, as discussed next.

\section{Thermodynamic Phase Diagrams}
\label{sec:tpd}

Despite tremendous efforts over several decades, while the exact phase 
diagram of even the simplest Hubbard model (which does not include 
the gauge fields or Zeeman fields) is still the subject of a hot debate, 
the mean-field phases and resultant phase diagrams of the mean-field 
Hubbard model are pretty much settled. To appreciate the effects of 
gauge fields, first we study Eq.~(\ref{eqn:ham}) with $\phi_\uparrow = \phi_\downarrow = 0$.

\subsection{No Gauge Fields: $\phi_\uparrow = \phi_\downarrow = 0$}
\label{sec:ngf}

\begin{figure}[htb]
\centerline{\scalebox{0.35}{\includegraphics{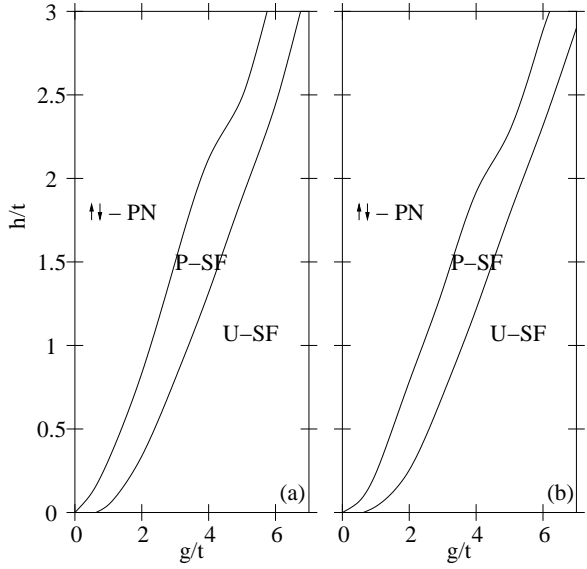}}}
\caption{\label{fig:01} (Color online) No gauge field: 
$\phi_\uparrow = \phi_\downarrow = 0$ case. 
The ground-state phase diagrams are shown for $\mu = 0$ in (a) and $\mu = -t$ in (b).
}
\end{figure}

Our results for this limit is presented in Fig.~\ref{fig:01}, where we set 
$\mu = 0$ in~\ref{fig:01}(a) corresponding to a half-filled lattice, and 
$\mu = -t$ in~\ref{fig:01}(b). 
We find that the phase diagrams are very similar, and depending on 
the particular value of $g$, there are two critical $h$ fields.
Since FFLO phase occupies a tiny parameter space near 
the normal phase boundary and only on the BCS side when 
$g/t \lesssim W$, we do not finely classify the character of P-SF 
phase in Fig.~\ref{fig:01} and throughout this paper. 
The U-SF phase, where $\Delta_i = \Delta_0$ for all $i$, turns 
into a P-SF beyond a first critical field $h_{c_1}$, and then the P-SF phase 
becomes an $\uparrow \downarrow$-$PN$ beyond a second critical field 
$h_{c_2} > h_{c_1}$. Our numerical results indicate that 
$h_{c_1} \sim |\Delta_0|$ where $|\Delta_0|$ is evaluated at $h = 0$ 
for the same parameters. 

In the strongly-interacting limit when $g \gg t$, it can be analytically shown 
for thermodynamic systems that $|\Delta_0| = (g/2-4t^2/g)\sqrt{n(2-n)}$, 
where $n = n_\uparrow + n_\downarrow$ is the total fermion filling. We
checked that this thermodynamic expression agrees very well with our 
finite-lattice results, as it gives $|\Delta_0| \approx 7.23 t$ 
for $\mu = 0$ or $n = 1$ and $|\Delta_0| \approx 7.18 t$ for $\mu = -t$ or 
$n \approx 0.875$ when $g = 15t$, while we find, respectively, 
$|\Delta_0| \approx 7.25 t$ and $|\Delta_0| \approx 7.19 t$ for the same 
parameters in our BdG calculations. In the weakly-interacting limit when 
$g$ is sufficiently small so that $\Delta_i \to 0$ for every $i$, we note that 
the system will be a $\downarrow$-vac for $h > 4t$ when $\mu = 0$ and 
for $h > 3t$ when $\mu =-t$. Next, we are ready to discuss the 
effects of balanced gauge fields.

\subsection{Balanced Gauge Fields: $\phi_\uparrow = \phi_\downarrow \ne 0$}
\label{sec:bgf}

\begin{figure}[htb]
\centerline{\scalebox{0.35}{\includegraphics{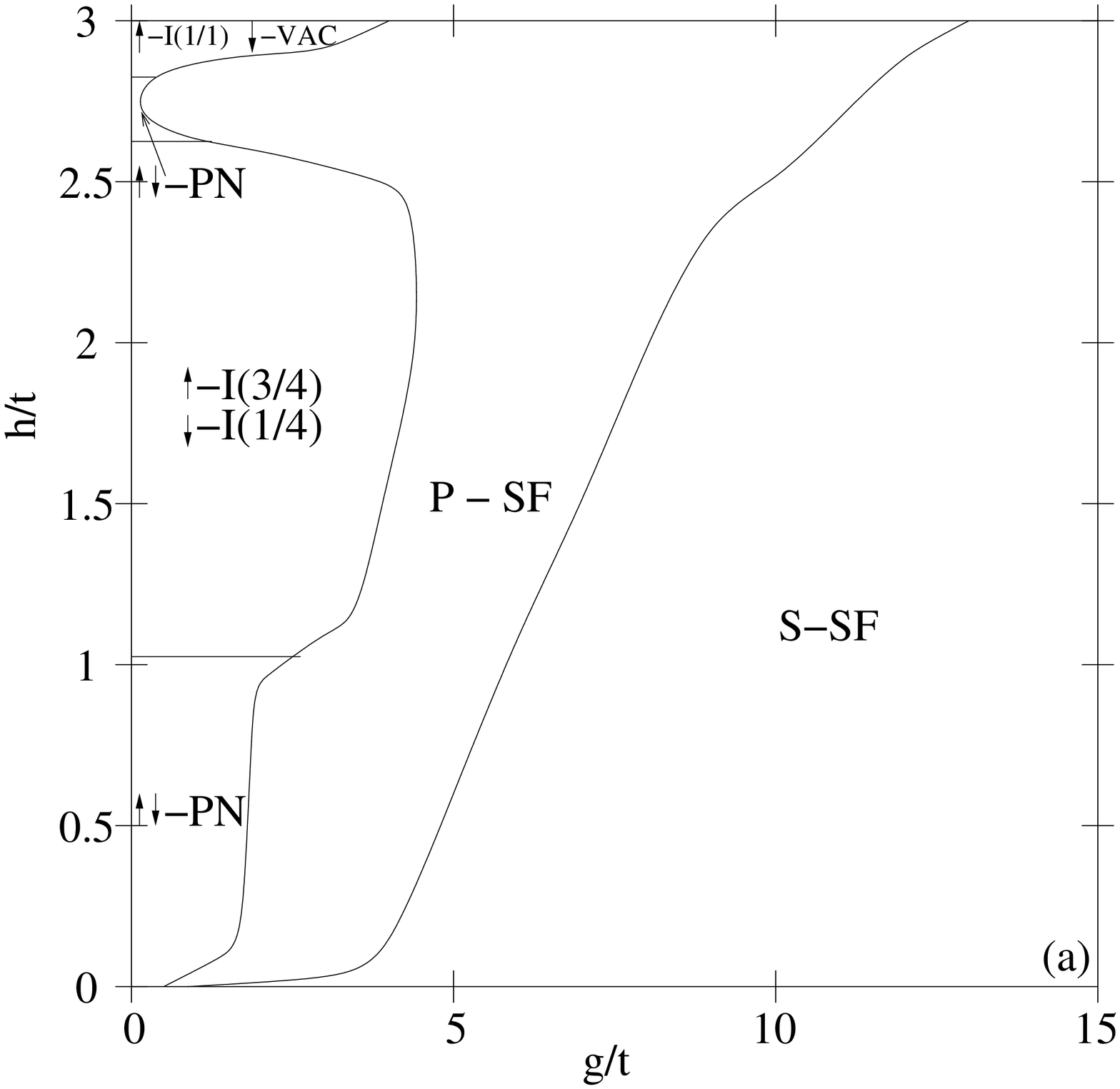}}}
\centerline{\scalebox{0.35}{\includegraphics{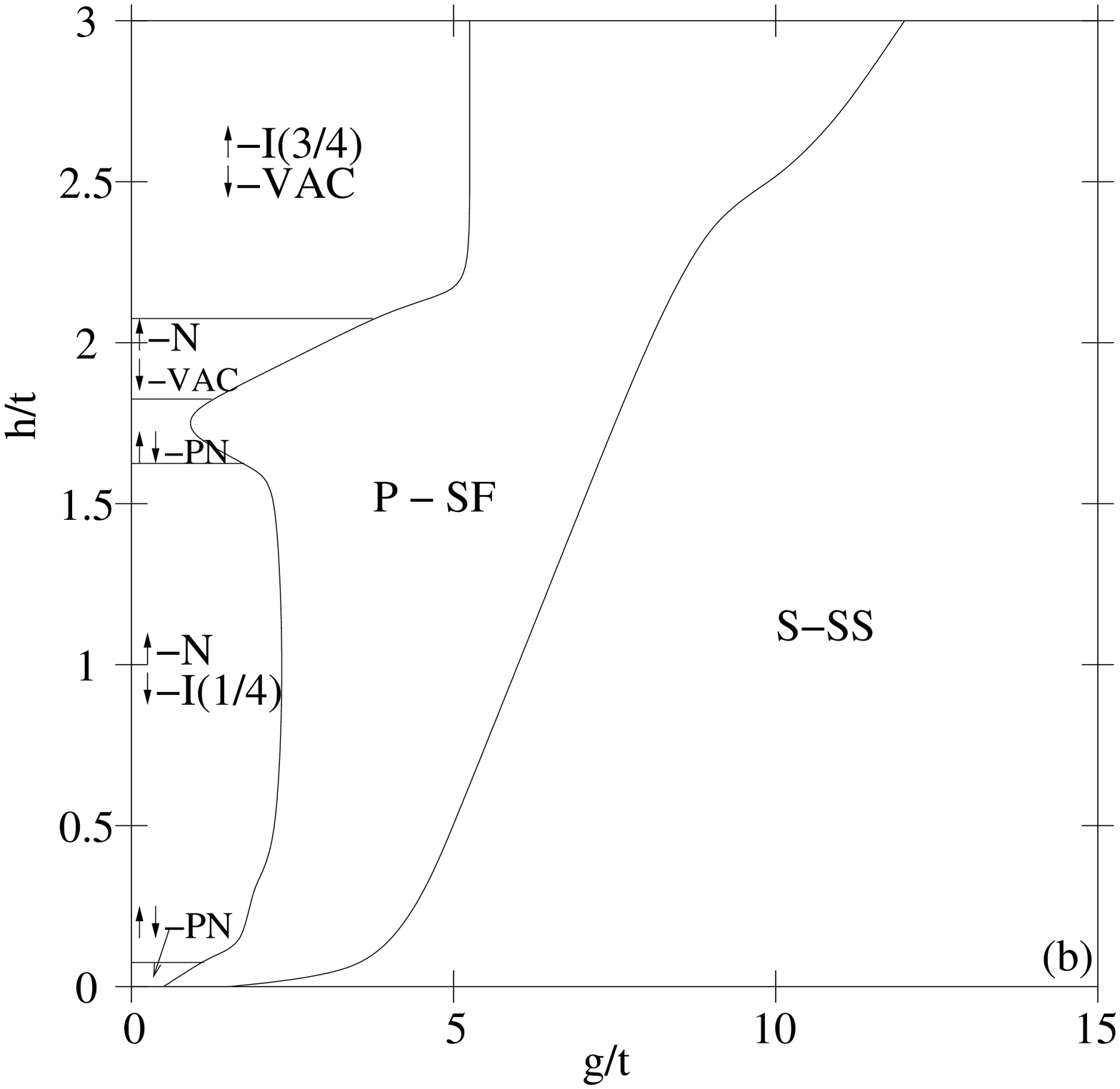}}}
\caption{\label{fig:14} (Color online) $\phi_\uparrow = \phi_\downarrow = 1/4$ case. 
The ground-state phase diagrams are shown for $\mu = 0$ in (a) and $\mu = -t$ in (b).
}
\end{figure}

In Fig.~\ref{fig:14}, we present the $\phi_\sigma = 1/4$ phase diagrams for 
$\mu = 0$ in~\ref{fig:14}(a) and $\mu = -t$ in~\ref{fig:14}(b). The $\mu = 0$ 
case is very special since it corresponds to a half-filled lattice with 
particle-hole symmetry, where $n_{i\uparrow} + n_{i\downarrow} = 1$ 
independently of $i$, no matter what the rest of the parameters are. 
In comparison to Fig.~\ref{fig:01}, the $\phi_\sigma = 1/4$ diagrams have 
much richer structure involving large regions of stripe-ordered phases. 
To understand the physical origin of the resultant phase diagrams and stripe 
order, next we discuss the analytically tractable high- and low-$h/g$ limits.

When $h/g$ is sufficiently high, we can directly read off the single-particle 
ground state of the $\sigma$ component from HB for any given 
$\phi_\sigma$. For $\phi_\sigma = 1/4$, the energy spectrum consists 
of 4 bands: the $\sigma$ component is 
a $\sigma$-vac for $\mu_\sigma \lesssim -2.83t$, 
a $\sigma$-$N$ for $-2.83t \lesssim \mu_\sigma \lesssim -2.61t$,
a $\sigma$-$I(1/4)$ for $-2.61t \lesssim \mu_\sigma \lesssim -1.082t$,
a $\sigma$-$N$ for $-1.082t \lesssim \mu_\sigma \lesssim 1.082t$,
a $\sigma$-$I(3/4)$ for $1.082t \lesssim \mu_\sigma \lesssim 2.61t$,
a $\sigma$-$N$ for $2.61t \lesssim \mu_\sigma \lesssim 2.83t$ and
a $\sigma$-$I(1/1)$ for $2.83t \lesssim \mu_\sigma$.
Using $\mu_\uparrow = \mu + h$ and $\mu_\downarrow = \mu - h$ in these
expressions, the high-$h/g$ structure of Fig.~\ref{fig:14} immediately follows.
As $h/g$ gets smaller, the single-particle I and N phases must pave the 
way to ordered many-body ones, as increasing the strength
of the pairing (attractive potential) energy eventually makes them energetically 
less favourable. For $\phi_\sigma = 0$, it is intuitively expected and 
numerically confirmed above that the $\uparrow \downarrow$-$PN$ to P-SF phase 
transition boundary $g(h_c)$ is a monotonic function of $h$, which is simply
because the non-interacting system has a very simple band structure 
with cosine dispersions. However, due to the presence of multiple bands, 
the transition boundary $g(h_c)$ becomes a complicated function of $h$ for 
finite $\phi_\sigma$. For instance, we find a sizeable hump in Fig.~\ref{fig:14}(a) 
around $h \approx 2.7t$ and another one in Fig.~\ref{fig:14}(b) around 
$h \approx 1.7t$, the peak locations of which coincide intuitively with the 
$\uparrow \downarrow$-$PN$ regions that are sandwiched between 
VAC and/or I. 

On the other hand, when $h/g$ is sufficiently small, the ground state is 
expected to be an ordered many-body phase with no polarisation. 
In sharp contrast to the $\phi_\sigma = 0$ case where U-SF is 
numerically confirmed above to be the ground state for any $\mu$, we show 
in Fig.~\ref{fig:14} that S-SF and S-SS are, respectively, stable for 
$\mu = 0$ and $\mu = -t$ when $\phi_\sigma = 1/4$. Note that since 
$\mu = 0$ corresponds to half filling for any $\phi_\sigma$, the 
unpolarized ground states necessarily have uniform fillings, i.e, 
$n_{i\uparrow} = n_{i\downarrow} = 1/2$ for every $i$. Therefore, 
in the low-$h/g$ limit, while only $|\Delta_i|$ is allowed to have spatial 
modulations in Fig.~\ref{fig:14}(a), both $|\Delta_i|$ and $n_{i \sigma}$ 
modulates in Fig.~\ref{fig:14}(b). 

\begin{figure}[htb]
\centerline{\scalebox{0.35}{\includegraphics{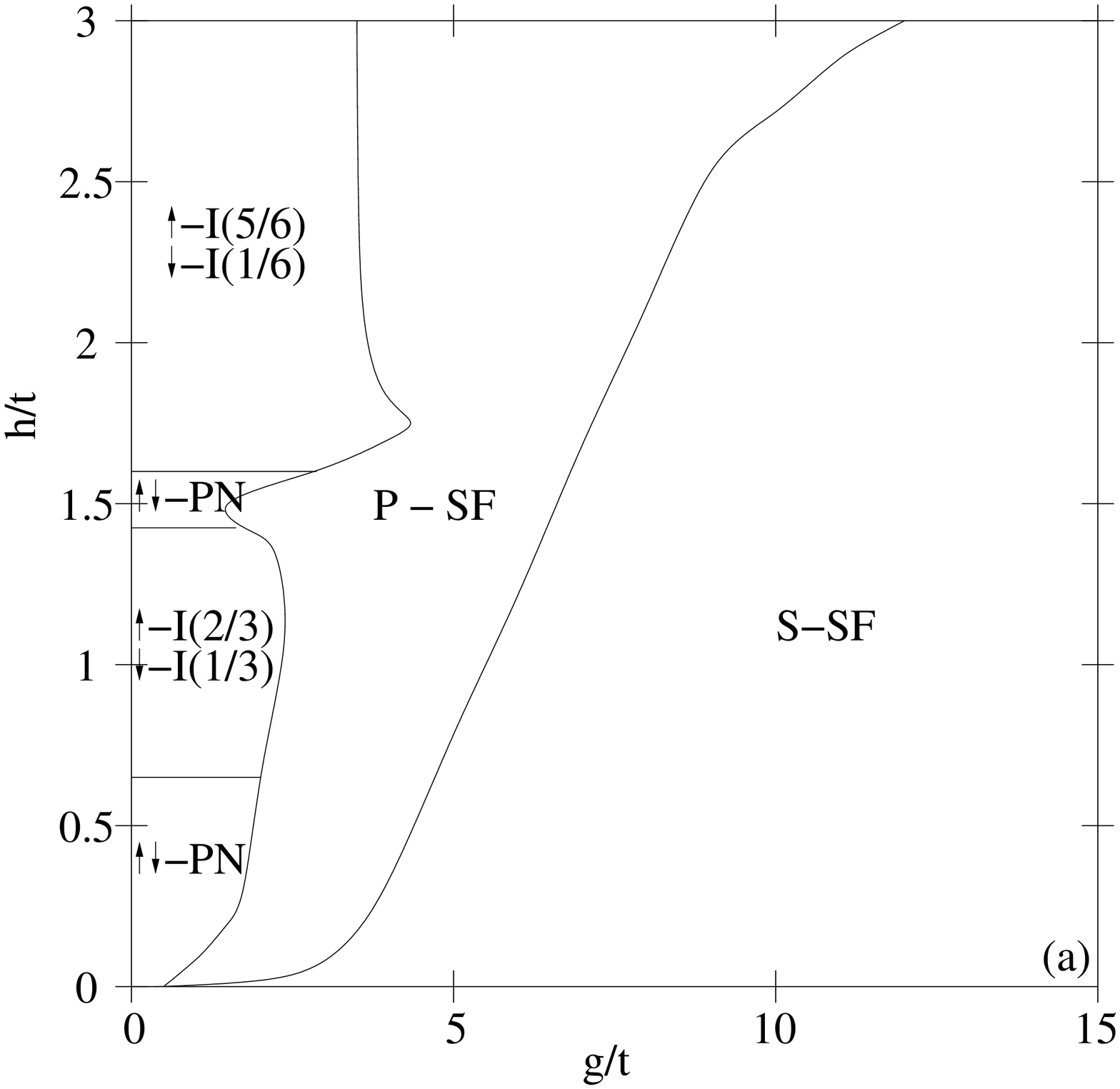}}}
\centerline{\scalebox{0.35}{\includegraphics{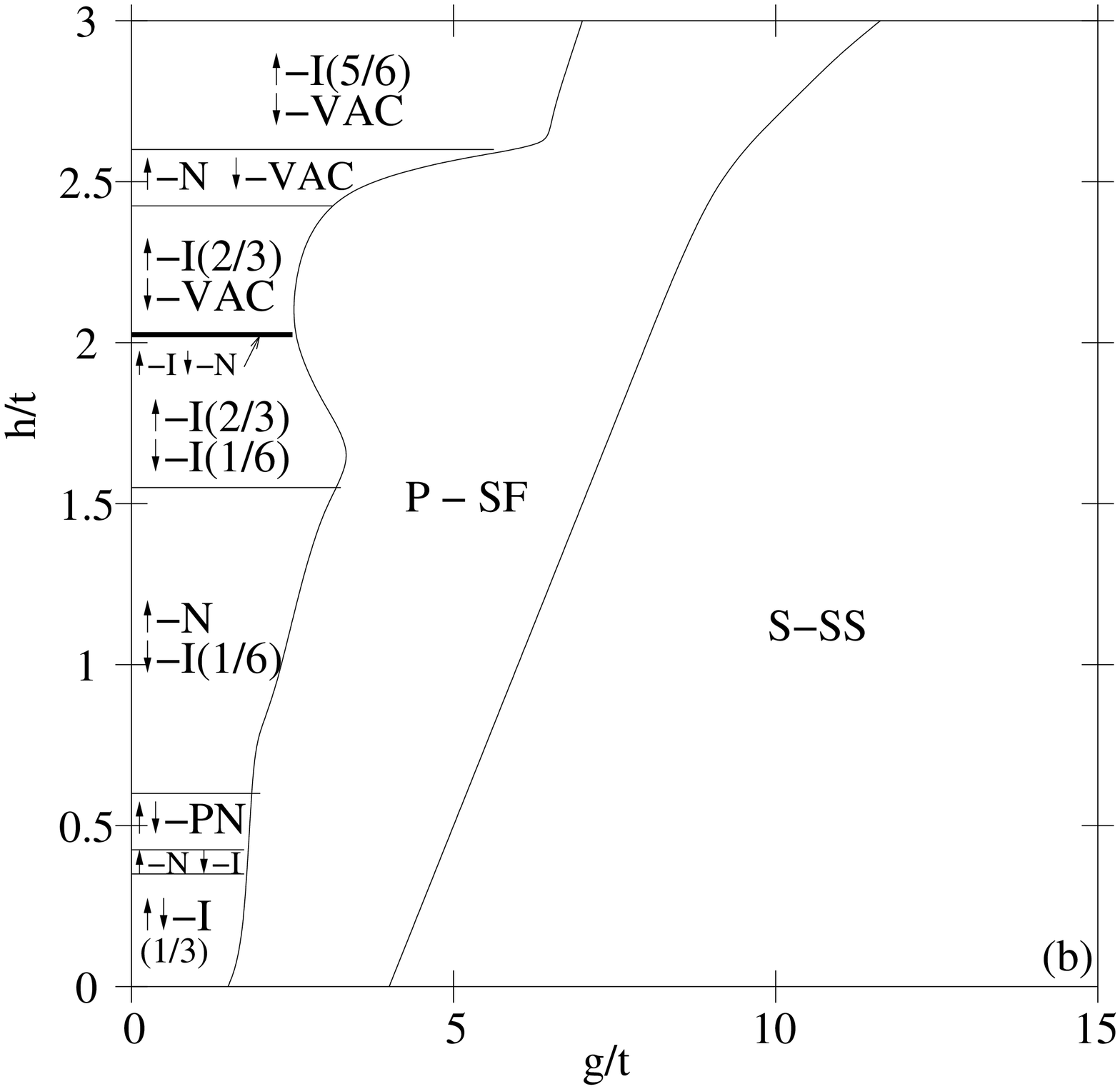}}}
\caption{\label{fig:16} (Color online) $\phi_\uparrow = \phi_\downarrow = 1/6$ case. 
The ground-state phase diagrams are shown for $\mu = 0$ in (a) and $\mu = -t$ in (b).
Note in (b) the presence of a sliver of $\downarrow$-$N$ region around $h \approx 2.076t$.  
}
\end{figure}

In comparison, the $\phi_\sigma = 1/6$ phase diagrams are shown 
in Fig.~\ref{fig:16}, and they are in many ways similar to the 
$\phi_\sigma = 1/4$ ones. 
The main difference is in the high-$h/g$ limit which again directly follows 
from HB. For $\phi_\sigma = 1/6$, the energy spectrum consists 
of 6 bands: the $\sigma$ component is 
a $\sigma$-vac for $\mu_\sigma \lesssim -3.076t$, 
a $\sigma$-$N$ for a narrow band around $\mu_\sigma \approx -3.076t$,
a $\sigma$-$I(1/6)$ for $-3.076t \lesssim \mu_\sigma \lesssim -1.59t$,
a $\sigma$-$N$ for $-1.59t \lesssim \mu_\sigma \lesssim -1.41t$,
a $\sigma$-$I(1/3)$ for $-1.41t \lesssim \mu_\sigma \lesssim -0.65t$,
a $\sigma$-$N$ for $-0.65t \lesssim \mu_\sigma \lesssim 0.65t$,
a $\sigma$-$I(2/3)$ for $0.65t \lesssim \mu_\sigma \lesssim 1.41t$,
a $\sigma$-$N$ for $1.41t \lesssim \mu_\sigma \lesssim 1.59t$,
a $\sigma$-$I(5/6)$ for $1.59t \lesssim \mu_\sigma \lesssim 3.076t$,
a $\sigma$-$N$ for a narrow band around $\mu_\sigma \approx 3.076t$
and $\sigma$-$I(1/1)$ for $3.076t \lesssim \mu_\sigma$.
As a consequence of this, we note in Fig.~\ref{fig:16}(b) that the system 
intuitively requires a finite threshold for $g/t$ even at $h = 0$, in order to 
develop any kind of many-body order. In addition, it is intriguing to see that 
the sliver of $\downarrow$-$N$ region that is sandwiched between 
$\downarrow$-vac and $\downarrow$-$I(1/6)$ around $h \approx 2.076t$ 
gives rise to a sizeable hump in Fig.~\ref{fig:16}(b). This is clearly a result of 
increased single-particle density of states.

Note in Figs.~\ref{fig:01}-\ref{fig:16} that the transition from an 
unpolarized to a polarized ordered phase occurs at a lower $h$ for 
any given $g$ as $\phi_\sigma$ is increased from $0$.
This is a consequence of smaller non-interacting energy bandwidths: 
as $\phi_\sigma$ increases from $0$ to $1/6$ to $1/4$ then $W$ 
shrinks from $8t$ to $6.15t$ to $5.65t$, making it possible to polarize 
the ground state with a smaller and smaller $h$. In Figs.~\ref{fig:14} 
and~\ref{fig:16}, the P-SF regions are dominated mainly by a 
phase that can be characterized by almost-striped PDW and SDW orders 
with some additional corrugations along the stripes that is caused by 
$h \ne 0$. For instance, when this phase is nearby to an insulating one, 
it generally has a very small SDW order in the background on top of 
a large and uniform polarisation.

\subsection{Imbalanced Gauge Fields: $\phi_\uparrow \ne \phi_\downarrow$}
\label{sec:igf}

As we argued in Secs.~\ref{sec:intro} and~\ref{sec:ham}, while independent 
control of the gauge fields $\phi_\uparrow$ and $\phi_\downarrow$ is 
not possible for conventional solid-state materials with real magnetic fields, 
such a control is plausible with neutral atomic systems. Motivated by this
exotic possibility, here we study two different limits.

\begin{figure}[htb]
\centerline{\scalebox{0.35}{\includegraphics{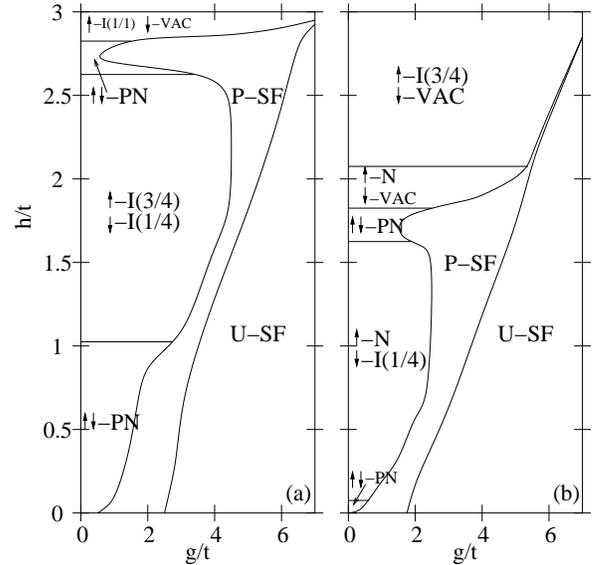}}}
\caption{\label{fig:14-14} (Color online) Time-reversal symmetric 
gauge fields: $\phi_\uparrow =  - \phi_\downarrow = 1/4$ case. 
The ground-state phase diagrams are shown for $\mu = 0$ in (a) and $\mu = -t$ in (b).
}
\end{figure}

As the first limit, we consider a pair of time-reversal symmetric gauge fields,
i.e, $\phi_\uparrow = - \phi_\downarrow$. 
For instance, $\phi_\uparrow = 1/4$ phase diagrams are shown in 
Fig.~\ref{fig:14-14}, where we set $\mu = 0$ in~\ref{fig:14-14}(a) and $\mu = -t$ 
in~\ref{fig:14-14}(b). Thanks to the time-reversal symmetry, even though 
the ground state is not a P-SF but an unpolarized SF at $h = 0$, it is 
not properly indicated in these figures for low $g/t$.
The general structures of the transition boundaries that
are seen in these phase diagrams are quite similar to the ones shown in 
Fig.~\ref{fig:14} for the $\phi_\uparrow = \phi_\downarrow = 1/4$ case. 
However, there is an important caveat in the dimer-BEC limit: 
the ground state becomes a U-SF for any $\mu$ as long as $h/g$ 
is sufficiently low. Given our analysis in Sec.~\ref{sec:dimer}, 
this is intuitively expected since the effective gauge field of Cooper pairs
vanish ($\phi_d = 0$) in the dimer-BEC limit as the gauge field of 
$\uparrow$ and $\downarrow$ fermions precisely cancel each other. 
In addition, the P-SF regions necessarily shrink here, since the 
U-SF to P-SF transition boundaries are expected to be close to the 
no-gauge-field ($\phi_\sigma = 0$) ones shown in Fig.~\ref{fig:01}.

\begin{figure}[htb]
\centerline{\scalebox{0.27}{\includegraphics{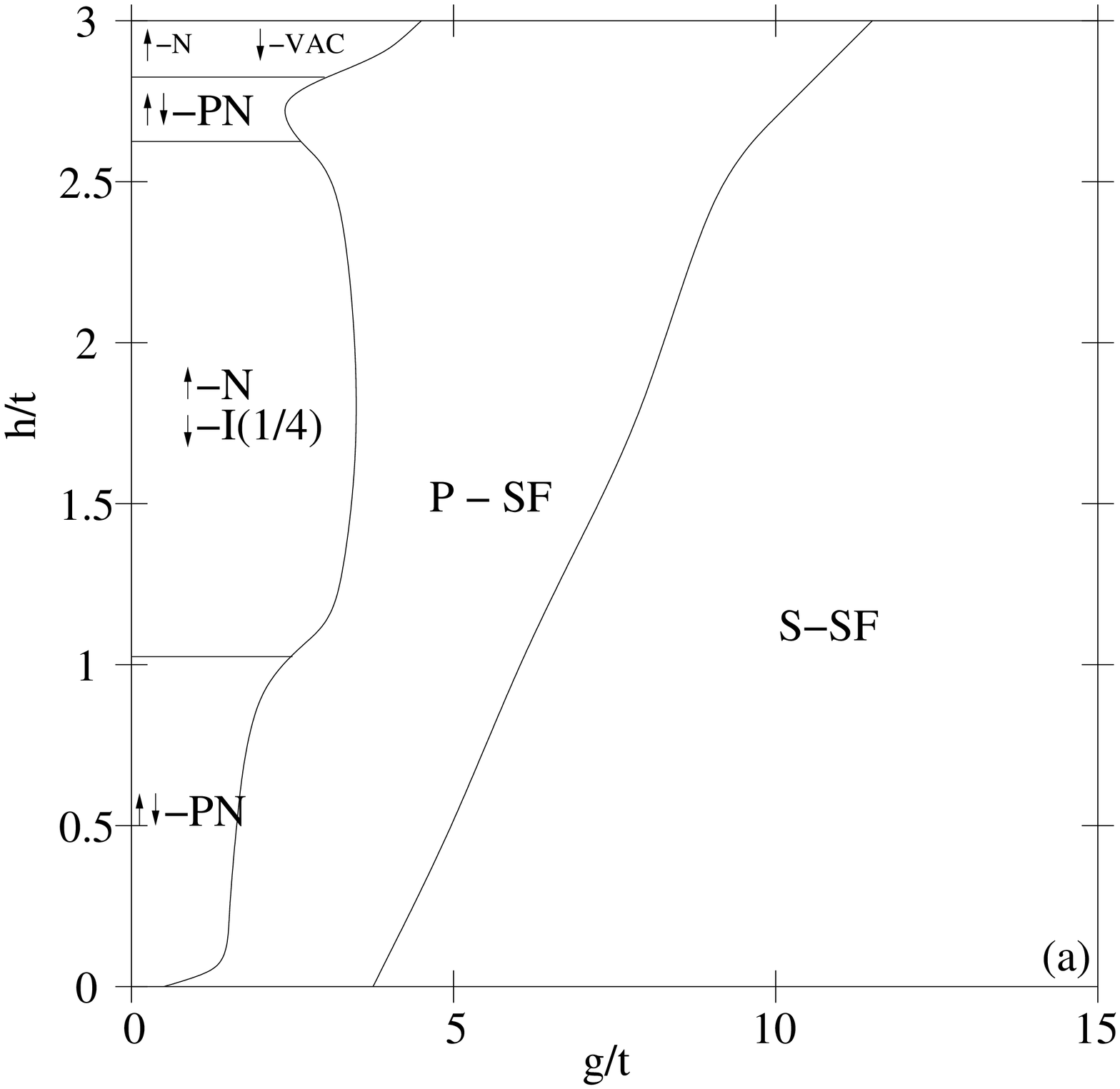}}}
\centerline{\scalebox{0.27}{\includegraphics{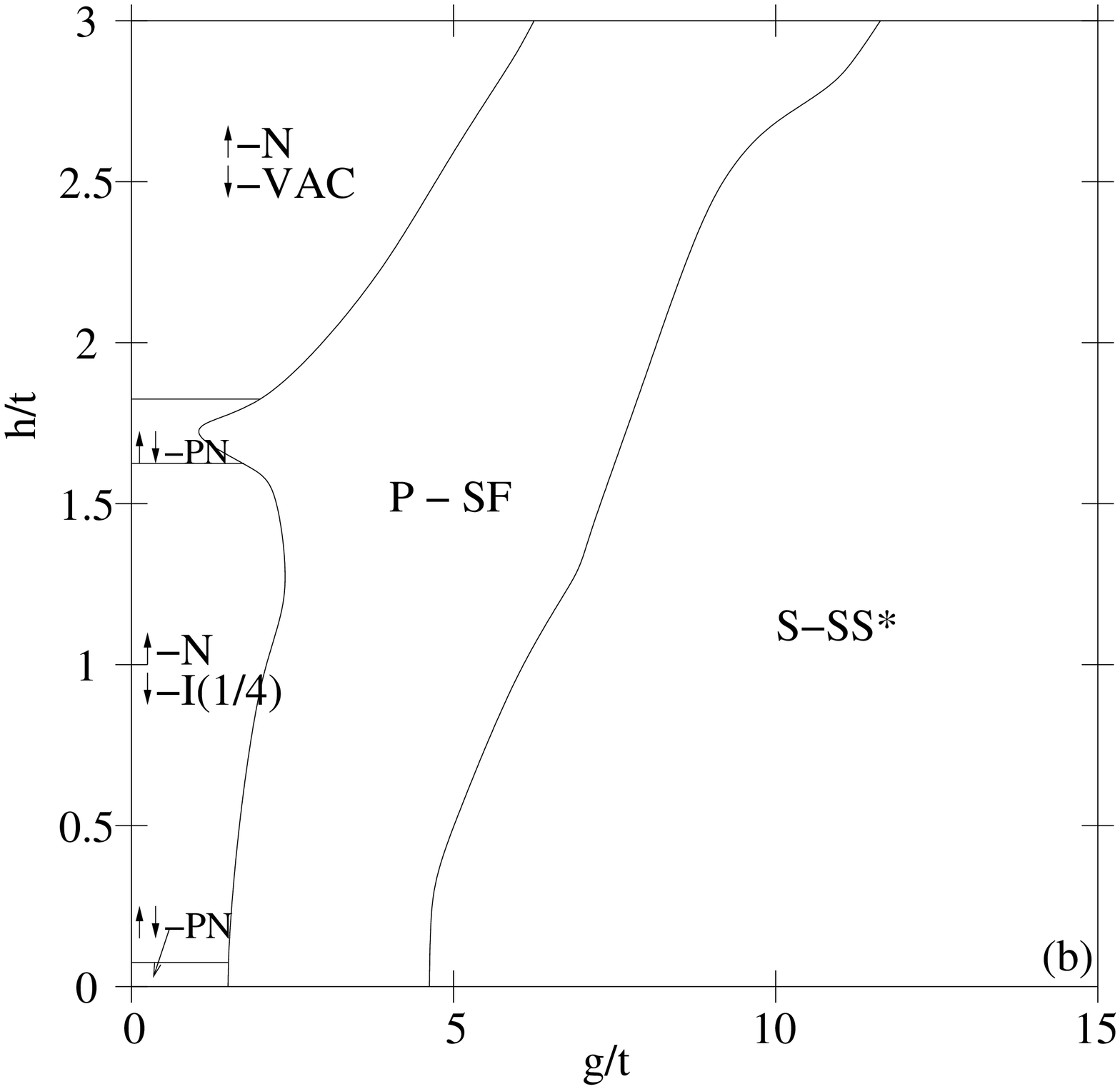}}}
\centerline{\scalebox{0.27}{\includegraphics{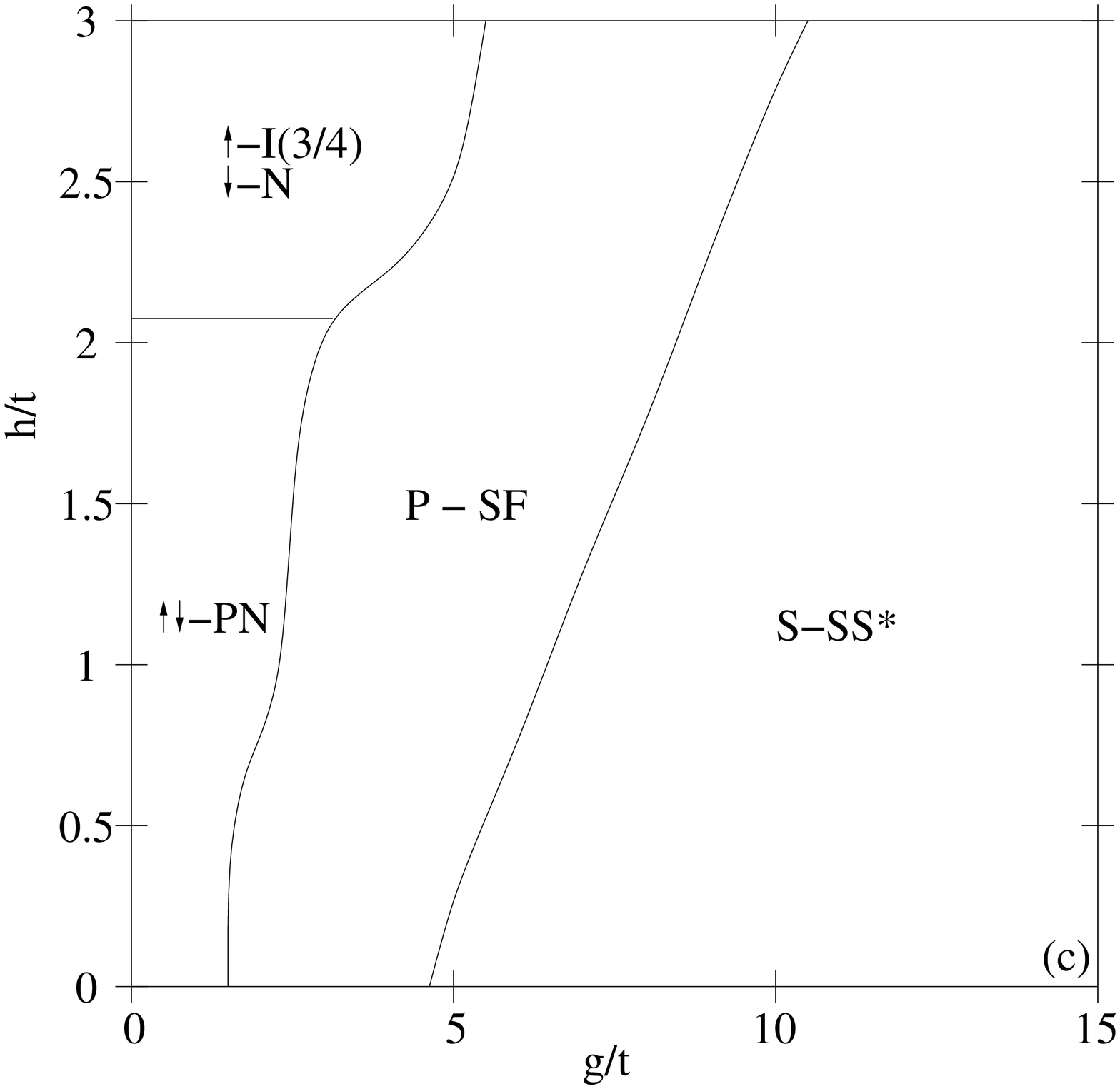}}}
\caption{\label{fig:0114} (Color online) Charged-uncharged 
mixtures of fermions: $\phi_\uparrow = 0$ and $\phi_\downarrow = 1/4$ 
case in (a-b) and $\phi_\uparrow = 1/4$ and $\phi_\downarrow = 0$ case in (c).
The ground-state phase diagrams are shown for $\mu = 0$ in (a) and $\mu = -t$ in (b-c).
Thanks to the particle-hole symmetry, the $\phi_\uparrow = 1/4$ and $\phi_\downarrow = 0$ 
phase diagram for $\mu = 0$ can easily be deduced from (a) via 
$\uparrow \to \downarrow$ and $\downarrow \to \uparrow$. 
}
\end{figure}

As the second limit, we set one of the gauge fields to zero, corresponding
effectively to a charged-uncharged mixture of two-component fermions. 
For instance, $(\phi_\uparrow = 0, \phi_\downarrow = 1/4)$ phase diagrams 
are shown in Fig.~\ref{fig:0114}, where we set $\mu = 0$ in ~\ref{fig:0114}(a) 
and $\mu = -t$ in~\ref{fig:0114}(b), and $(\phi_\uparrow = 1/4, \phi_\downarrow = 0)$ 
diagram is shown in Fig.~\ref{fig:0114}(c) where we set $\mu = -t$.
Thanks to the particle-hole symmetry around half-filling, 
$(\phi_\uparrow = 1/4, \phi_\downarrow = 0)$ phase diagram for $\mu = 0$ 
can easily be deduced from~\ref{fig:0114}(a) via $\uparrow \to \downarrow$ and 
$\downarrow \to \uparrow$, and therefore, it is not shown. 
Since this symmetry also prevents polarisation at $h = 0$, even though the 
ground state is not a P-SF but an unpolarized non-uniform (but non-striped) SF for 
weak $g/t$, this is not properly indicated in Fig.~\ref{fig:0114}(a). 
However, the imbalance between gauge fields causes P-SF in 
Figs.~\ref{fig:0114}(b) and~\ref{fig:0114}(c) even at $h = 0$. Similar to the
analysis given in Sec.~\ref{sec:bgf}, the high- and low-$h/g$ limits can be 
directly read off from HB and effective dimer-BEC descriptions, respectively, 
with again an important caveat in the dimer-BEC limit: the ground state 
becomes a S-SS* for $\mu \ne 0$ as long as $h/g$ is sufficiently low. 
As shown in Fig.~\ref{fig:stripes}(c), in addition to the coexisting striped-PDW 
and -CDW orders, S-SS*has an additional sign-changing striped-SDW 
order driven solely by $\phi_\uparrow \ne \phi_\downarrow$. 
Note also that if $(\phi_\uparrow \ne 0, \phi_\downarrow = 0)$ then all of the 
coexisting orders of S-SS* phase are periodic along the $y$ direction 
with periodicity $q_d = q_\uparrow$ since $\phi_d = \phi_\uparrow$.

\subsection{Stripe Order vs. FFLO Modulations}
\label{sec:sm}

It is clearly the cooperation between g, $\phi_\uparrow$ and $\phi_\downarrow$ 
that is responsible for the broken spatial symmetry and 
appearance of stripe order, causing much more prominent stripes for 
intermediate $g$ at a given $h$. 
The stripe order is a direct result of HB: for a given $\phi_\sigma$, 
the spectrum consists of $q_\sigma$-bands in the 1st magnetic Brillouin 
zone within which each $\mathbf{k}$ state is $q_\sigma$-fold degenerate. 
Therefore, when $g \ne 0$, not only intra- and inter-band pairings but also 
pairings with both 0 and a set of non-zero center-of-mass momenta are 
allowed~\cite{zhai10, wei12}, leading to a non-uniform $|\Delta_i|$ with 
spatially-periodic modulations, e.g, a PDW order~\cite{agterberg08}. 
The directions of center-of-mass momenta determine the direction 
of modulations, making it gauge dependent, e.g, $y$ direction 
in Fig.~\ref{fig:stripes}. When the striped-PDW order is sufficiently large, 
it drives an additional striped-CDW order in the total fermion filling, 
giving rise to striped-SS phases. 

We emphasise that the instabilities towards stripe-ordered phases discussed 
in this paper are driven by the gauge fields, and they may formally not be identified 
with the FFLO phase which is driven by the Zeeman field and is characterized 
by cosine-like sign-changing $|\Delta_i|$ oscillations 
along a spontaneously-chosen direction~\cite{FF64, LO65, casalbuoni04}. 
In addition, while the periods of
our striped-PDW, -CDW and -SDW orders are always given by $q_d$, 
the period of FFLO modulations is determined by the mismatch $h$ between 
$\uparrow$ and $\downarrow$ Fermi surfaces. 
For instance, when $\phi_\uparrow = \phi_\downarrow = p/q$, the stripes 
have a spatial period of $q$ or $q/2$ lattice sites, depending on whether 
$q$ is odd or even. Lastly, while our striped phases survive even in the 
extreme dimer-BEC limit ($g/t \gg 1$) for a large parameter space, the FFLO 
modulations survive not only in the BCS limit but also for a tiny parameter 
space nearby the P-SF to N transition boundary.

\section{Confined Atomic Systems}
\label{sec:cas}

Having explored the ground states and phase diagrams of thermodynamic 
systems, here we study confined systems and comment on the likelihood of 
observing stripe-ordered phases by loading neutral atomic Fermi gases 
on laser-induced optical lattices under laser-generated artificial gauge fields.
For this purpose, we consider a harmonically-confined $51 \ell \times 51 \ell$ 
square lattice with an isotropic trapping potential $V_i = \alpha |\mathbf{r_i}|^2$
centered at the origin, where $\alpha = 0.01t/\ell^2$ is its strength and 
$\mathbf{r_i} \equiv (i_x, i_y)$ is the position of site $i$.

\subsection{Effects of Harmonic Confinement}
\label{sec:harmonic}

The local ground states of trapped systems can be reliably inferred 
through the so-called local-density approximation, where the local 
density of the system is mapped to that of a thermodynamic one 
with the same density. This description is known to be very 
accurate for large systems that are trapped in slowly-varying 
potentials. For our model Hamiltonian, due to the energy gaps of 
HB and the Pauli exclusion principle, one expects the so-called 
wedding-cake structures in $n_{i\uparrow}$ and $n_{i\downarrow}$ 
profiles of non-interacting fermions at $T = 0$, 
where the number of mini-gaps determines the number of
spatially-flat $n_{i\sigma}$ regions for a given $\phi_\sigma$. 
Thus, wedding-cake structures consist of a number of insulating 
regions that are sandwiched between normal regions.
However, since the majority of these mini-gaps are very small
compared to $t$, finite $g$ and/or finite $T$ quickly smear out 
the flat regions, making their detection nearly impossible. 
In sharp contrast, here we show that the broken spatial symmetry 
and stripe orders persist at intermediate and strong interactions, 
providing a viable knob for the experimental probe of the fractal 
structure of HB.

\begin{figure}[htb]
\centerline{\scalebox{0.75}{\includegraphics{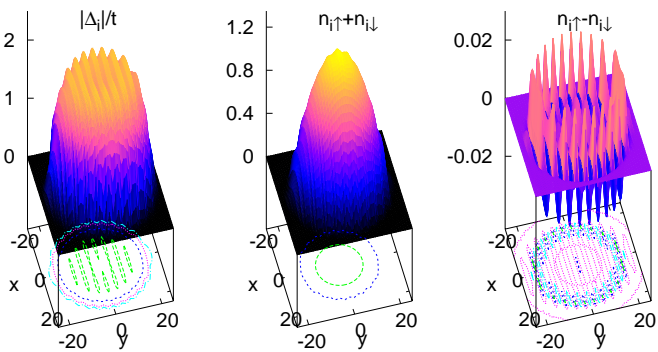}}}
\caption{\label{fig:g5-1401-noHS} (Color online) 
The trap profiles are shown for $\phi_\uparrow = 1/4$, $\phi_\downarrow = 0$, 
$\mu = t$, $h = 0$ and $g = 5t$. Here, $(x,y)$ are in units of $\ell$.
}
\end{figure}

In Fig.~\ref{fig:g5-1401-noHS}, we illustrate a typical self-consistent solution 
for a trapped system when $\phi_\uparrow = 1/4$, $\phi_\downarrow = 0$, 
$\mu = t$, $h = 0$ and $g = 5t$. The total numbers of $\sigma$ fermions are 
approximately given by $N_\uparrow = N_\downarrow \approx 464$. 
While the remnants of the so-called wedding-cake structure, 
i.e., spatially-flat $n_{i\uparrow}$ regions around integer multiples of 
$1/4$ fillings, are hardly recognisable, a large PDW order is clearly visible. 
Given the phase diagrams discussed in Sec.~\ref{sec:igf}, both CDW 
and SDW orders are expected to be weak around half-filling, since 
$n_{i\uparrow} + n_{i \downarrow} \gtrsim 1$ near the center of the trap 
for this particular set of data.

\begin{figure}[htb]
\centerline{\scalebox{0.7}{\includegraphics{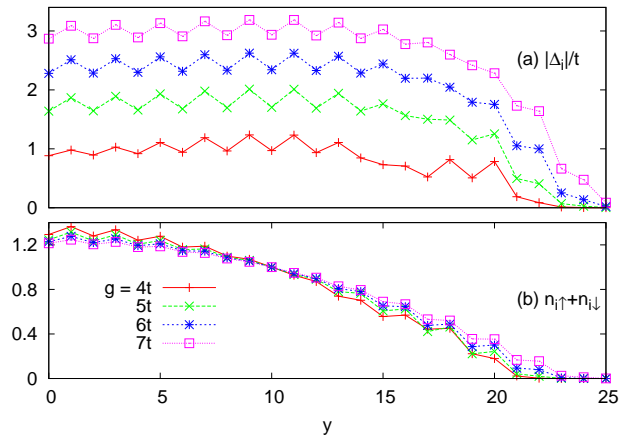}}}
\caption{\label{fig:1414-noHS} (Color online)
The trap profiles are shown for $x = 0$ cuts along the $y$ direction
which is in units of $\ell$.
Here, $\phi_\uparrow = \phi_\downarrow = 1/4$, $\mu = t$ and $h = 0$,
and therefore, the system is locally unpolarized at every $i$.
}
\end{figure}

\begin{figure}[htb]
\centerline{\scalebox{0.7}{\includegraphics{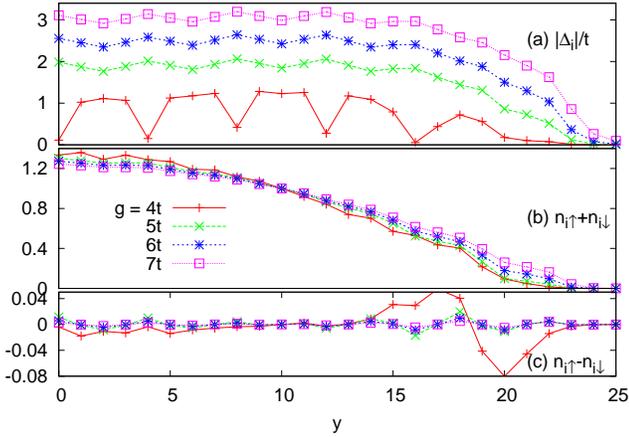}}}
\caption{\label{fig:1401-noHS} (Color online)  
The trap profiles are shown for $x = 0$ cuts along the $y$ direction
which is in units of $\ell$.
Here, $\phi_\uparrow = 1/4$, $\phi_\downarrow = 0$, $\mu = t$ and $h = 0$.
}
\end{figure}

It is easier to visualise and present such trap profiles for a cut along the 
$y$ direction at a particular $x$ value. For instance, we show $x = 0$ 
cuts in Figs.~\ref{fig:1414-noHS} and~\ref{fig:1401-noHS}, 
where $\phi_\uparrow = 1/4$, $\mu = t$ and $h = 0$ in both figures, but 
$\phi_\downarrow = 1/4$ and $\phi_\downarrow = 0$, respectively.
While the local ground states are always unpolarized in 
Fig.~\ref{fig:1414-noHS} where $n_{i\uparrow} = n_{i\downarrow}$ for every $i$, 
the imbalance between $\phi_\uparrow$ and $\phi_\downarrow$ causes 
small but visible SDW orders in Fig.~\ref{fig:1401-noHS}.
We note that $\phi_\uparrow \ne \phi_\downarrow$ may also cause a 
global polarisation, i.e, $N_\uparrow \ne N_\downarrow$, for 
weak $g$, however, this polarisation must gradually disappear towards 
the dimer-BEC limit. For instance, as $g/t$ increases to $(4,5,6,7)$, 
while $N_\uparrow = N_\downarrow$ is approximately given by 
$(454, 468, 491, 519)$ in Fig.~\ref{fig:1414-noHS}, $N_\uparrow$ 
and $N_\downarrow$ are given, respectively, by $(456, 464, 489,518)$ 
and $(448, 464, 489,518)$ in Fig.~\ref{fig:1401-noHS}.

These figures show that the CDW and SDW orders tend to be more 
prominent for intermediate $g$ as long as the system is away from
half-filling. This is quite intuitive since the appearance of a PDW order 
breaks the spatial symmetry of the system at the first place. 
The spatial periods are, respectively, given by 2 and 4 sites 
in Figs.~\ref{fig:1414-noHS} and~\ref{fig:1401-noHS}, and these findings 
are in agreement with our analysis given in Sec.~\ref{sec:dimer}.
In addition, since the relative stripes eventually fade away towards 
the dimer-BEC limit, the trap profiles slowly recover the usual 
(no-gauge-field) results in both figures. It is also pleasing to see that the 
valleys of the PDW and CDW orders and peaks of the SDW order 
coincide when they coexist. These results suggest that observation of 
PDW, CDW and SDW features as a function of magnetic flux may furnish 
clearest and direct evidence for the existence of multiple band structure, 
and hence indirectly for the fractal HB, in trapped atomic systems.

\subsection{Effects of Hartree Shifts}
\label{sec:hs}

Since most of our phases have either coexisting CDW and/or SDW 
orders, our phase diagrams may not be convenient to generate more 
accurate phase diagrams by including the Hartree terms 
via a simple shift in $\mu_{i \sigma}$. 
However, we still neglected these shifts in our diagrams for their 
numerical as well as analytical simplicity.  For instance, including 
these shifts in the self-consistency Eqs.~(\ref{eqn:bdg})-(\ref{eqn:ndo}) 
not only requires about an order of magnitude more iterations 
to converge, but also it complicates our current intuition 
making it more difficult to extract the relation between HB and the 
non-monotonic dependences of some of the phase boundaries.
Note that since Hartree shifts have no role in driving the stripe-ordered 
phases, which is particularly clear in the dimer-BEC limit where 
$\mu_{i \sigma}$ do not explicitly play any role in our analysis, their 
inclusion is expected to change some of the transition boundaries 
without much effect on the stability of phases. 
Furthermore, since the mean-field theory provides only a qualitative 
description of the phase diagrams and the accuracy of our results can be 
somewhat improved by including these shifts, one still needs to 
go beyond this approximation for experimentally more relevant diagrams. 
Therefore, even though Hartree shifts are neglected in Sec.~\ref{sec:tpd}, 
our results may already pave the way to qualitative understanding of
the exact ground states of the attractive Hofstadter-Hubbard model.

\begin{figure}[htb]
\centerline{\scalebox{0.7}{\includegraphics{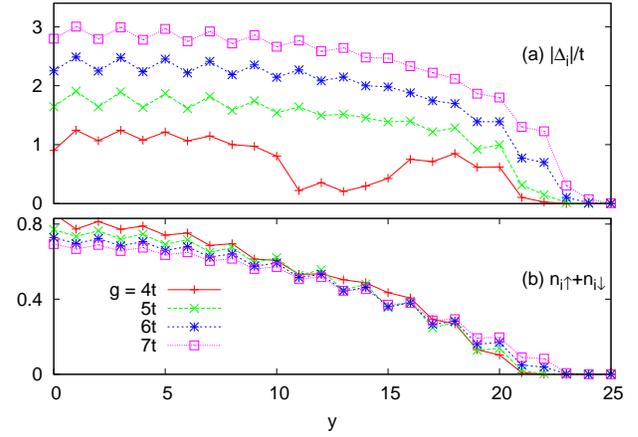}}}
\caption{\label{fig:1414-HS} (Color online)  
The trap profiles are shown for $x = 0$ cuts along the $y$ direction
which is in units of $\ell$.
Here, $\phi_\uparrow = \phi_\downarrow = 1/4$, $\mu = t$ and $h = 0$,
i.e., same as Fig.~\ref{fig:1414-noHS} with the Hartree shifts included.
}
\end{figure}

\begin{figure}[htb]
\centerline{\scalebox{0.7}{\includegraphics{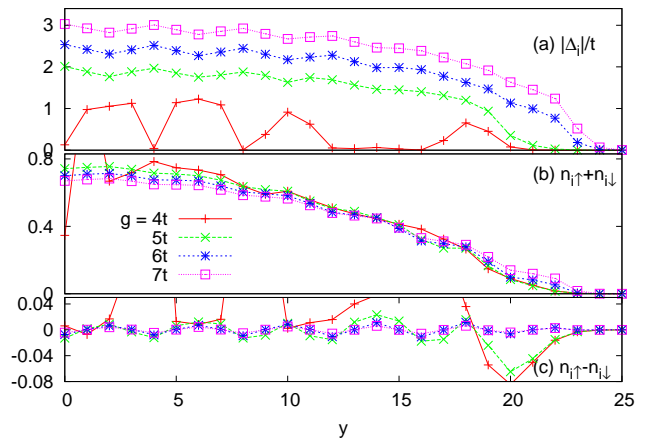}}}
\caption{\label{fig:1401-HS} (Color online)  
The trap profiles are shown for $x = 0$ cuts along the $y$ direction
which is in units of $\ell$.
Here, $\phi_\uparrow = 1/4$, $\phi_\downarrow = 0$, $\mu = t$ and $h = 0$,
i.e., same as Fig.~\ref{fig:1401-noHS} with the Hartree shifts included.
}
\end{figure}

To illustrate these points, the Hartree-shifted trap profiles are shown 
in Figs.~\ref{fig:1414-HS} and~\ref{fig:1401-HS} for the parameters 
of Figs.~\ref{fig:1414-noHS} and~\ref{fig:1401-noHS}, respectively.
Comparing these figures show that while the inclusion of the 
Hartree shifts does not have much effect on $|\Delta_i|$ for these 
particular sets of data (thanks to the particle-hole symmetry around 
half-filling), it affects the total filling quite a bit. 
For instance, as $g/t$ increases to $(4,5,6,7)$, while $N_\uparrow = N_\downarrow$ 
is approximately given by $(288, 279, 279, 286)$ in Fig.~\ref{fig:1414-HS}, 
$N_\uparrow$ and $N_\downarrow$ are given, respectively, 
by $(298, 278, 278, 285)$ and $(276, 282, 278, 285)$ in Fig.~\ref{fig:1401-HS}.
However, the visibility of the striped-PDW and -CDW orders remain 
largely the same in both cases. 
In addition, we note that the remnants of the wedding-cake structures, 
i.e., spatially-flat $n_{i\uparrow} + n_{i\downarrow}$ regions 
around $1/2$ fillings, are almost recognisable in Figs.~\ref{fig:1414-HS}(a) 
and~\ref{fig:1401-HS}(a) when $g = 4t$ or less (not shown). 
While the non-interacting $\uparrow$ fermions are insulating at 
$1/4$ filling in both figures, the non-interacting $\downarrow$ fermions 
are insulating (normal) in Fig.~\ref{fig:1414-HS} (\ref{fig:1401-HS}). 
Thus, these insulating regions leave their traces as distinct 
$|\Delta_i|$ dips in both figures near $y = 13 \ell$ when $g$ is 
sufficiently weak. 

Given these numerical illustrations, it is clear that our phase diagrams 
already shed some light on a new stripe mechanism in the dimer-BEC limit, 
showing that the fate of stripe-ordered SF and SS phases 
are not affected by the Hartree terms. 
Having discussed the effects of confinement potentials, we are ready to 
end the paper with a briery summary of our conclusions and an outlook.

\section{Conclusions}
\label{sec:conc}

Our mean-field results for the attractive single-band Hofstadter-Hubbard 
model on a square lattice are as follows. In the presence of a Zeeman field 
$h$, in addition to the intriguing phase transition boundaries between 
the N/I/VAC and SF phase, we found a number of distinct 
many-body phases which can 
be characterized with respect to their coexisting striped-PDW, -CDW 
and -SDW orders. Even at $h = 0$, we reached four important conclusions.
First, we numerically found an unpolarized striped-superfluid phase 
(S-SF) in a large parameter space. Unlike the conventional FFLO phase 
which is driven by $h$, our S-SF is driven only by the gauge fields. 
Second, we numerically found an unpolarized striped-supersolid phase 
(S-SS) in a large parameter space. Unlike the conventional SS phase 
which is yet to be observed and is driven either by long-range 
(e.g., nearest-neighbor) interactions or the presence of a second species 
(e.g., Bose-Fermi or Bose-Bose mixtures), our S-SS is again driven only by 
the gauge fields. Third, we also found a locally polarized but globally unpolarized 
striped-SS phase (S-SS*) when the gauge fields are imbalanced.
Lastly, we provided analytical insights on the microscopic origins of these 
stripe-ordered phases, suggesting a new physical mechanism that gives 
rise to FFLO-like SF and SS phases in the dimer BEC limit.

The importance of these results can be highlighted as follows. 
First, spatially-modulated SF and SS phases are both of high interest not 
only to the atomic physics community but also to the condensed-matter, 
nuclear and elementary-particle physics communities. 
Second, the unusual appearance of the stripe order is very exotic and 
fundamentally important by itself, because the connection between the
striped-charge order that is observed in copper-oxide materials and the 
formation of high-Tc superconductivity has been the subject of a long 
debate in the literature. Even though our work offers no direct relation to 
cuprate superconductors, understanding stripe-ordered phases in the cold-atom 
context may still prove to be beneficial for the high-Tc community.
Third, the existence of stripe-ordered phases is not an artefact of our 
mean-field BdG description, since they are analytically motivated in the 
dimer-BEC limit. Therefore, we highly encourage further research in this 
direction with different lattice geometries, gauge fields, etc., 
in particular the beyond mean-field ones.

\begin{acknowledgments}
This work is supported by T\"{U}B$\dot{\mathrm{I}}$TAK Grant No. 1001-114F232,
and the author thanks Dr. R. O. Umucal{\i}lar for discussions.
\end{acknowledgments}

\appendix

\section{Dimer-BEC Limit in the Symmetric Gauge}

It may be important to remark here that the analysis given in Sec.~\ref{sec:dimer} 
depends on the particular artificial gauge field that is simulated in a cold-atom 
experiment. Next, we use a rotationally-invariant symmetric gauge for the 
vector potential, i.e, $\mathbf{A_\sigma}(\mathbf{r}) \equiv B_\sigma (-y, x,0)/2$, 
and analyse the spatial structure of the order parameter $|\Delta_i|$ in the 
dimer-BEC limit, which clearly reveals this dependence.

\begin{figure}[htb]
\centerline{\scalebox{0.7}{\includegraphics{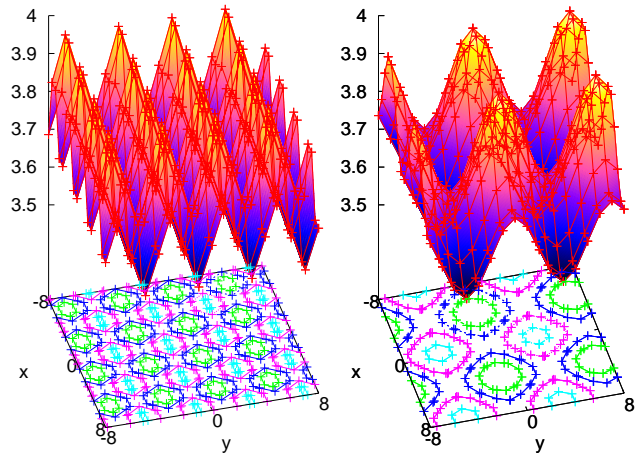}}}
\caption{\label{fig:cstripes} (Color online)
Typical $|\Delta_i|/t$ profiles are shown for the checkerboard-like stripes, 
where $\phi_\uparrow = \phi_\downarrow = 1/4$ in the left and $1/8$ in the right 
figure as determined by the symmetric gauge. 
Here, $(x,y)$ are in units of $\ell$, and we set $\mu = 0$, $h = 0$ and 
$g = 8t$ in both figures.
}
\end{figure}

Typical $|\Delta_i|$ profiles are illustrated in Fig.~\ref{fig:cstripes} for $\phi_\sigma = 1/4$ 
and $1/8$, i.e. $\phi_d = \phi_\uparrow + \phi_\downarrow \equiv p_d/q_d$ is $1/2$ 
and $1/4$, respectively.
Since the $C_4$ symmetry of the square lattice is preserved in this gauge,
the stripes are checkerboard-like in the $(x,y)$ plane, such that 
generalisation of Eq.~(\ref{eqn:fit}) to
\begin{align}
|\Delta_i| = |\Delta_0| + |\Delta_1| \left[ 2 - \cos(\pi \phi_d i_x/\ell) - \cos(\pi \phi_d i_y/\ell) \right]
\label{eqn:fits}
\end{align}
fits very well with all of our numerical results in the dimer-BEC limit. 
Here, the uniform contribution $|\Delta_0| = (g/2-4t^2/g)\sqrt{n(2-n)}$ is 
determined by the total average filling $n$ with
$
\mu = (g/2-8t^2/g) (n-1)
$ 
~\cite{iskin-lattice}, 
$|\Delta_1| \approx t^2/g$ for $\mu \approx 0$,
and $\mathbf{r_i} \equiv (i_x, i_y)$ is the position of site $i$.
We again note that this analytical expression is consistent with our expectation 
that while the dimer-BEC order parameter in principle has contributions from all 
degenerate momenta
$
\Psi_{id} = \sum_{\mathbf{k_d}} c_{\mathbf{k_d}} e^{i \mathbf{k_d} \cdot \mathbf{r_i}},
$
where $c_{\mathbf{k_d}}  = |c_{\mathbf{k_d}}| e^{i \vartheta_{\mathbf{k_d}}}$ are complex 
variational parameters, only the lowest order $\mathbf{k_d} = \{ (0,0); \pi \phi_d (f_x,f_y) / \ell \}$ 
terms contribute when $g/t$ is sufficiently large. This is because dimer bosons are 
fermion pairs and the number of ways of creating them with 
$\mathbf{k_d} = \mathbf{k_{\uparrow}} + \mathbf{k_{\downarrow}}$ 
momentum depends on $f_x$, $f_y$, $\phi_\uparrow$ and $\phi_\downarrow$, 
and higher $\mathbf{k_d}$ states contribute less and less, forming again 
a perturbative series. 

In addition, moving towards the BCS side, the second-order corrections to 
Eq.~(\ref{eqn:fits}) can be shown to be
$+ |\Delta_2| [\cos(2\pi \phi_d i_x/\ell) + \cos(2\pi \phi_d i_y/\ell) ]$ and
$+ |\Delta_2'| \cos(\pi \phi_d i_x/\ell) \cos(\pi \phi_d i_y/\ell)$
for any $q_d$.
Since these terms are in- (out-of-$\pi$-) phase with the zeroth (first) order term, 
they tend to create dimples at the peaks arised from the first order one,
suggesting that $\vartheta_{\mathbf{k_d}} - \vartheta_{\mathbf{0}} = \pi (f_x+f_y)$,
i.e., the form of $|\Psi_{id}|$ again coincides with $|\Delta_i|$ under these conditions.
Finally, we note that both our numerical results as well as the analytical fit 
Eq.~(\ref{eqn:fits}) clearly show that modulations of $|\Delta_i|$ have a spatial 
period of $2q_d$ lattice sites along both $x$ and $y$ directions, as expected 
for the resultant dimers in the symmetric gauge.

\end{document}